\documentclass[aps,prl,twocolumn,showpacs,groupedaddress,longbibliography]{revtex4-1}

\usepackage{graphics}
\usepackage[pdftex]{graphicx}
\usepackage[pdftex]{epsfig}
\usepackage{epstopdf}

\usepackage{epsfig}
\usepackage{amsmath}
\usepackage{amssymb}
\usepackage{amsfonts}

\usepackage{color}
\usepackage{ulem}

\newcommand{\bra}[1]{\ensuremath{\langle#1|}}
\newcommand{\ket}[1]{\ensuremath{|{#1}\rangle}}
\newcommand{\braket}[2]{ \langle #1 | #2 \rangle }

\newcommand{\beginsupplement}{%
        \setcounter{table}{0}
        \renewcommand{\thetable}{S\arabic{table}}%
        \setcounter{figure}{0}
        \renewcommand{\thefigure}{S\arabic{figure}}%
        \setcounter{equation}{0}
        \renewcommand{\theequation}{S\arabic{equation}}
        \setcounter{section}{0}
        \renewcommand{\thesection}{S\Roman{section}}
     }

\begin{document}

\title{Dressed photon-orbital states in a quantum dot: Inter-valley spin resonance}

\author{P. Scarlino}
\author{E. Kawakami}
\author{T. Jullien}
\affiliation{Kavli Institute of Nanoscience, TU Delft, Lorentzweg 1, 2628 CJ Delft, The Netherlands}
\author{D. R. Ward}
\author{D. E. Savage}
\author{M. G. Lagally}
\author{Mark Friesen}
\author{S. N. Coppersmith}
\author{M. A. Eriksson}
\affiliation{University of Wisconsin-Madison, Madison, WI 53706, USA}
\author{L. M. K. Vandersypen}
\affiliation{Kavli Institute of Nanoscience, TU Delft, Lorentzweg 1, 2628 CJ Delft, The Netherlands}
\affiliation{Components Research, Intel Corporation, 2501 NW 29th Ave, Hillsboro, OR 97124, USA}

\date{\today}
\vskip1.5truecm

\begin{abstract}
The valley degree of freedom is intrinsic to spin qubits in Si/SiGe quantum dots. It has been viewed alternately as a hazard, especially when the lowest valley-orbit splitting is small compared to the thermal energy, or as an asset, most prominently in proposals to use the valley degree of freedom itself as a qubit. Here we present experiments in which microwave electric field driving induces transitions between both valley-orbit and spin states. We show that this system is highly nonlinear and can be understood through the use of dressed photon-orbital states, enabling a unified understanding of the six microwave resonance lines we observe. Some of these resonances are inter-valley spin transitions that arise from a non-adiabatic process in which both the valley and the spin degree of freedom are excited simultaneously. For these transitions, involving a change in valley-orbit state, we find a tenfold increase in sensitivity to electric fields and electrical noise compared to pure spin transitions, strongly reducing the phase coherence when changes in valley-orbit index are incurred. In contrast to this non-adiabtaic transition, the pure spin transitions, whether arising from harmonic or subharmonic generation, are shown to be adiabatic in the orbital sector. The non-linearity of the system is most strikingly manifest in the observation of a dynamical anti-crossing between a spin-flip, inter-valley transition and a three-photon transition enabled by the strong nonlinearity we find in this seemly simple system.
\end{abstract}
\maketitle

A spin-1/2 particle is the canonical two-level quantum system. Its energy level structure is extremely simple, consisting of just the spin-up and spin-down levels. Therefore, when performing spectroscopy on an elementary spin-1/2 particle such as an electron spin, only a single resonance is expected corresponding to the energy separation between the two levels.

\begin{figure*}[ht!]
\includegraphics[width=14cm]{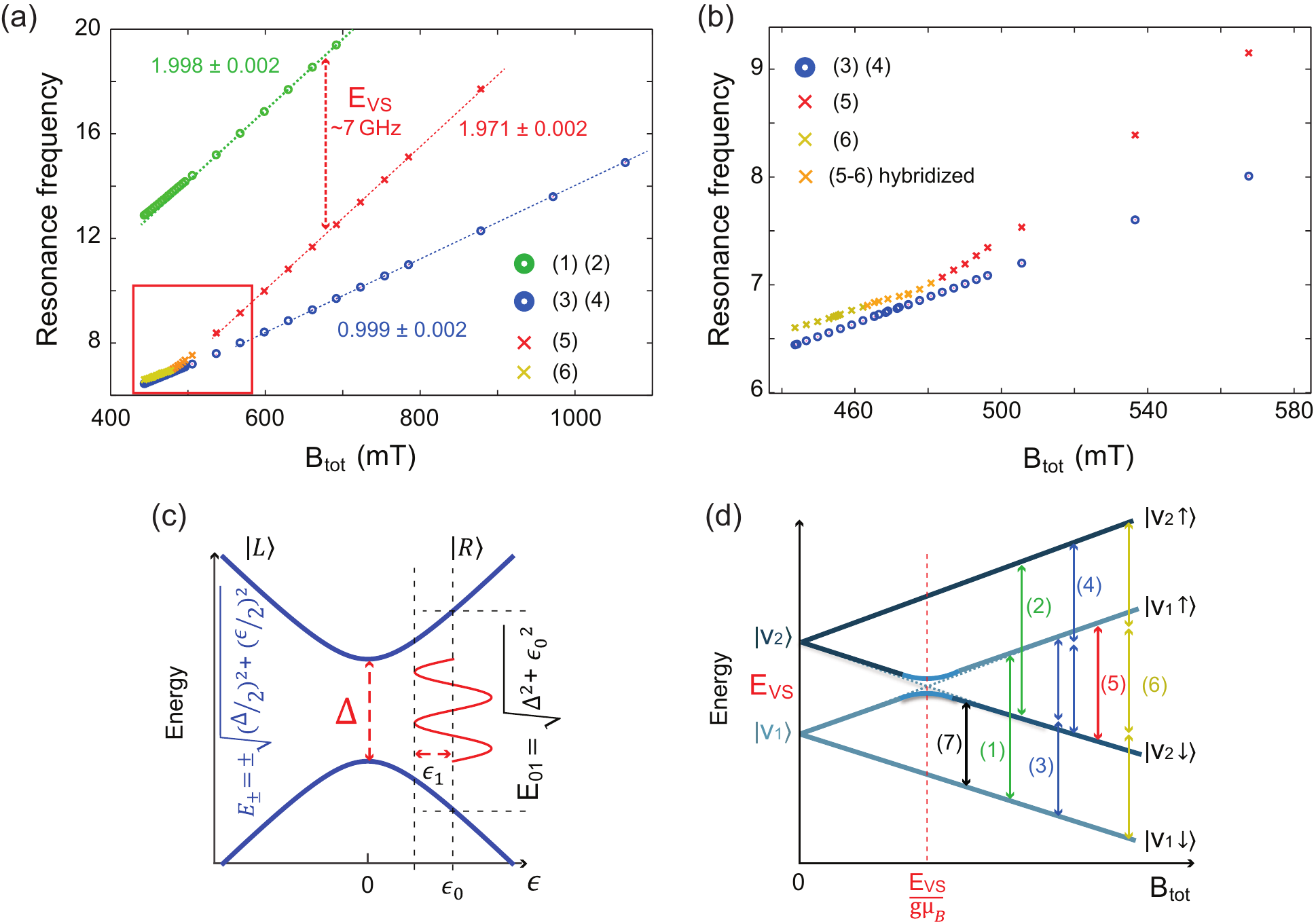}
\caption{(a) Multiple resonance frequencies as a function of the external magnetic field, observed for a single electron spin confined in a gate-defined Si/SiGe quantum dot, driven by low-power microwave excitation applied to one of the quantum dot gates. Resonances (1) and (2) are indistinguishable on this scale, as are resonances (3) and (4). On the horizontal axis, we plot $B_\text{tot}=\sqrt{(B^x_\text{ext}+B_\|)^2+B^2_\perp}$, with $B_{\|} \sim - 120$ mT and $B_{\perp} \sim 50$ mT the estimated components of the stray magnetic field from the micromagnet [magnetized, in this measurement, along the $x$ direction (see Fig.~S3)]. (b) Zoom-in of the region indicated by a red box in panel (a). (c) Schematic energy level diagram of a generic two-level quantum system described by Eq.~(\ref{eq:realH}), as a function of detuning $\epsilon$ and with a harmonic driving of amplitude $\epsilon_1$ around the central value $\epsilon_0$ with energy eigenvalues $E_\pm$ and energy splitting $E_{01}=E_+ - E_-$.
(d) The four energy levels considered in this work as a function of $B_\text{tot}$, in the absence of photonic dressing, comprised of two valley-orbit states ($\ket{V_{1,2}}$) and two spin states ($\ket{\downarrow,\uparrow}$). $E_\text{VS}$ represents the valley-orbit energy splitting. The vertical arrows labelled 1-6 correspond to  6 processes observed in our simulations and to excitations observed in the experiment, with the same labeling scheme as panels (a) and (b). (1) and (2)
correspond to single-photon spin flips. (3) and (4) correspond to two-photon spin flips. (5) corresponds to a combined spin flip,
valley-orbit excitation, and single-photon absorption. (6) is also a combined spin-flip, valley-orbit excitation, between different states than (5), and is a three-photon process. (7) is a nonadiabatic valley-orbit excitation; it is observed in simulations but not in the experiments, because $E_\text{VS} \sim k_B T_\text{el}$, and therefore readout is ineffective.
}
\label{fig:Fig1b} 
\end{figure*}

Recent measurements have shown that the spectroscopic response of a single electron spin in a quantum dot can be much more complex than this simple picture suggests. This is particularly true when using electric-dipole spin resonance, where an oscillating electric field couples to the spin via spin-orbit coupling \cite{Nowack2007}. First, due to non-linearities in the response to oscillating driving fields, subharmonics can be observed \cite{Laird2009,Pei2012,Laird2013,Stehlik2014,Nadj-Perge2012,Forster2015}, and the non-linear response can even be exploited for driving coherent spin rotations \cite{Scarlino2015}. Second, due to spin-orbit coupling, the exact electron spin resonance frequency in a given magnetic field depends on the orbital the electron occupies \cite{Khaetskii2001}. In silicon or germanium quantum dots, the conduction band valley is an additional degree of freedom \cite{Friesen2006, Friesen2010, Goswami2007, Zwanenburg2013, Rancic2016}, and the electron spin resonance frequency should depend on the valley state as well \cite{Rancic2016,Yang2013, Hao2014,Kawakami2014,Veldhorst2014,Veldhorst2015}. As a result, when valley or orbital energy splittings are comparable to or smaller than the thermal energy, thermal occupation of the respective levels leads to the observation of multiple closely spaced spin resonance frequencies \cite{Kawakami2014}.

The picture becomes even richer when considering transitions in which not only the spin state but also the orbital quantum number changes. Such phenomena are common in optically active dots \cite{Warburton2013}, but have been observed also in electrostatically defined (double) quantum dots in the form of relaxation from spin triplet to spin singlet states \cite{Fujisawa2002,Johnson2005} and spin-flip photon-assisted tunneling \cite{Schreiber2011,Braakman2014}. However, that work is all in semiconductor quantum dots with no valley degree of freedom, and the degree to which valleys -often treated as weakly coupled to each other and orbital states-couple to each other to enable microwave-driven transitions that change spin has not been explored.
Furthermore, the investigation of resonant transitions involving the valley degree of freedom is very important in the context of the new qubit architecture recently proposed for Si quantum dots \cite{Culcer2012}, based on the valley degree of freedom to encode and process quantum information.

Here, we report transitions where both the spin and valley-orbit state flip in a Si/SiGe quantum dot. We demonstrate that we can Stark shift the transitions, and we compare the sensitivity to electric fields to the case of pure spin transitions, including the impact on phase coherence. We find that the valley-orbit coupling strongly affects the coherence properties of the inter-valley spin resonances.
We show that a theory incorporating a driven four-level system comprised of two valley-orbit and two spin states subject to strong ac driving provides a consistent description of these transitions, as well as all the previously reported transitions for this system. This theory also explains the observation of a dynamical level repulsion, which can be understood effectively and compactly using a dressed-state formalism.

\section{Device and spectroscopic measurements}

The device used for this experiment has been described in \cite{Kawakami2014} (see Fig.~S3). It is based on an undoped Si/SiGe heterostructure with two layers of electrostatic gates. Two accumulation gates are used to induce a two-dimensional electron gas (2DEG) in a 12 nm wide Si quantum well 37 nm below the surface and a set of depletion gates is used to form a single quantum dot in the 2DEG, and a charge sensor next to this dot. The dot is tuned so it is occupied by just one electron. Two micromagnets placed on top of the accumulation gates produce a local magnetic field gradient. The sample is attached to the mixing chamber of a dilution refrigerator with base temperature $\sim$25 mK and an electron temperature estimated from transport measurements of $\sim$150 mK. For the present gate voltage configuration, the valley splitting, $E_\text{VS}$, is comparable to the thermal energy, $k_BT_\text{el}$.

Microwave excitation applied to one of the gates oscillates the electron wave function back and forth in the dot, roughly along the $x$ axis (Fig.~S3). Because of the local magnetic field gradient $dB_\perp/dx\sim$0.3 mT/nm \cite{Kawakami2014}, where $B_\perp$ is the component of the micromagnet field gradient perpendicular to the static magnetic field $B_\text{ext}$, the electron is then subject to an oscillating magnetic field \cite{Tokura2006,Pioro-Ladriere2008} and electron spin transitions can be induced when the excitation is resonant with the spin splitting. The spin-up probability $P_\uparrow$ in response to the microwave excitation is measured by repeated single-shot cycles (see Sec.~S.III.A of the Supplemental Material for details). 
The initialization and read-out procedures require a Zeeman splitting exceeding $k_B T_\text{el}$, which here restricts us to working at $B_\text{tot}>450$ mT.

When varying the applied microwave frequency and external magnetic field, we observe six distinct resonance peaks~\footnote{Here, we present measurements realized for different magnetic field orientations. The components of the external magnetic field are reported in each figure. The specific orientation of the external magnetic field does not play any special role. It is our understanding that the results presented in this work are independent from the specific magnetic field orientation.} [see Fig.~\ref{fig:Fig1b}(a)]. The two resonances labeled (1) and (2), not resolved on this scale, are two intra-valley spin resonances, one for each of the two lowest-lying valley states that are thermally occupied \cite{Kawakami2014}. They exhibit a $T_2^\ast \sim 1$ $\mu$s and Rabi frequencies of order MHz. The two resonances labeled (3) and (4), similarly not resolved, arise from second harmonic driving of the two intra-valley spin flip transitions. These transitions too can be driven coherently, with Rabi frequencies comparable to those for the fundamental harmonic, as we reported in \cite{Scarlino2015}.

We focus here on the resonances labeled (5) and (6) in Fig. 1, which have not been discussed before. The frequency of resonance (5), $f_0^{(5)}$, is $\sim$7 GHz lower than the fundamental intra-valley spin resonance frequencies, $f_0^{(1)}$ and  $f_0^{(2)}$. From the magnetic field dependence measured above 500 mT, we extract a $g$-factor of about 1.971 $\pm$ 0.002, close to but different from the $g$-factors for resonances (1) and (2) (1.99 \cite{Kawakami2014}). The line width [Fig.~\ref{fig:Fig5}(a) inset and Fig.~S4] is almost ten times larger than that for the intra-valley resonances, giving a correspondingly shorter $T_2^\ast$ of around 100 ns. Around 500 mT, resonance (5) changes in a way reminiscent of level repulsions and transitions into resonance (6) [see Fig.~\ref{fig:Fig1b}(b)]. Without the change in slope, resonance (5) would have crossed resonances (3) and (4); however, the latter do not show any sign of level repulsion and continue their linear dependence on magnetic field.

We interpret these puzzling observations starting with Fig.~\ref{fig:Fig1b}(d). Two sets of Zeeman split levels are seen, separated by the energy of the first excited valley-orbit state. The green (1) [(2)] and two blue (3) [(4)] arrows show driving of spin transitions via the fundamental and second harmonic respectively, for the valley-orbit ground [excited] state. We identify resonance (5) with the transition indicated with the red arrow in which both spin and valley(-orbit) flip. It has the same field dependence as resonance (1), but (above 500 mT) it is offset from resonance (1) by a fixed amount, which as we can see from Fig.~\ref{fig:Fig1b}(d), is a measure of the valley-orbit splitting, $E_\text{VS}$.
Resonance (6) is a three photon process in which both spin and valley-orbit states flip. As we discuss below, hybridization between (5) and (6) is possible at magnetic fields $B_\text{tot} = \frac{2 E_\text{VS}}{g \mu_B}$, for which the photons shown by the corresponding arrows in Fig.~\ref{fig:Fig1b}(d) have the same energy.

\section{Model}

We now introduce a simple model Hamiltonian that can be used to understand the observed spectroscopic response. This model explains the presence of both the first and second harmonic driven spin resonance as well as the observed inter-valley spin resonance. We show that resonances such as those observed in Figs.~\ref{fig:Fig1b}(a) and \ref{fig:Fig1b}(b) are generic features of a strongly driven four-level system composed of two orbital levels and two spin levels
in which there is a coupling between the orbital levels, such as a tunnel coupling.
For our case, it is natural to associate the orbital levels with two different valley-orbit states (see Sec.~S.I of the Supplemental Material for details).

\begin{figure*}[ht!]
\centering
\includegraphics[width=\linewidth]{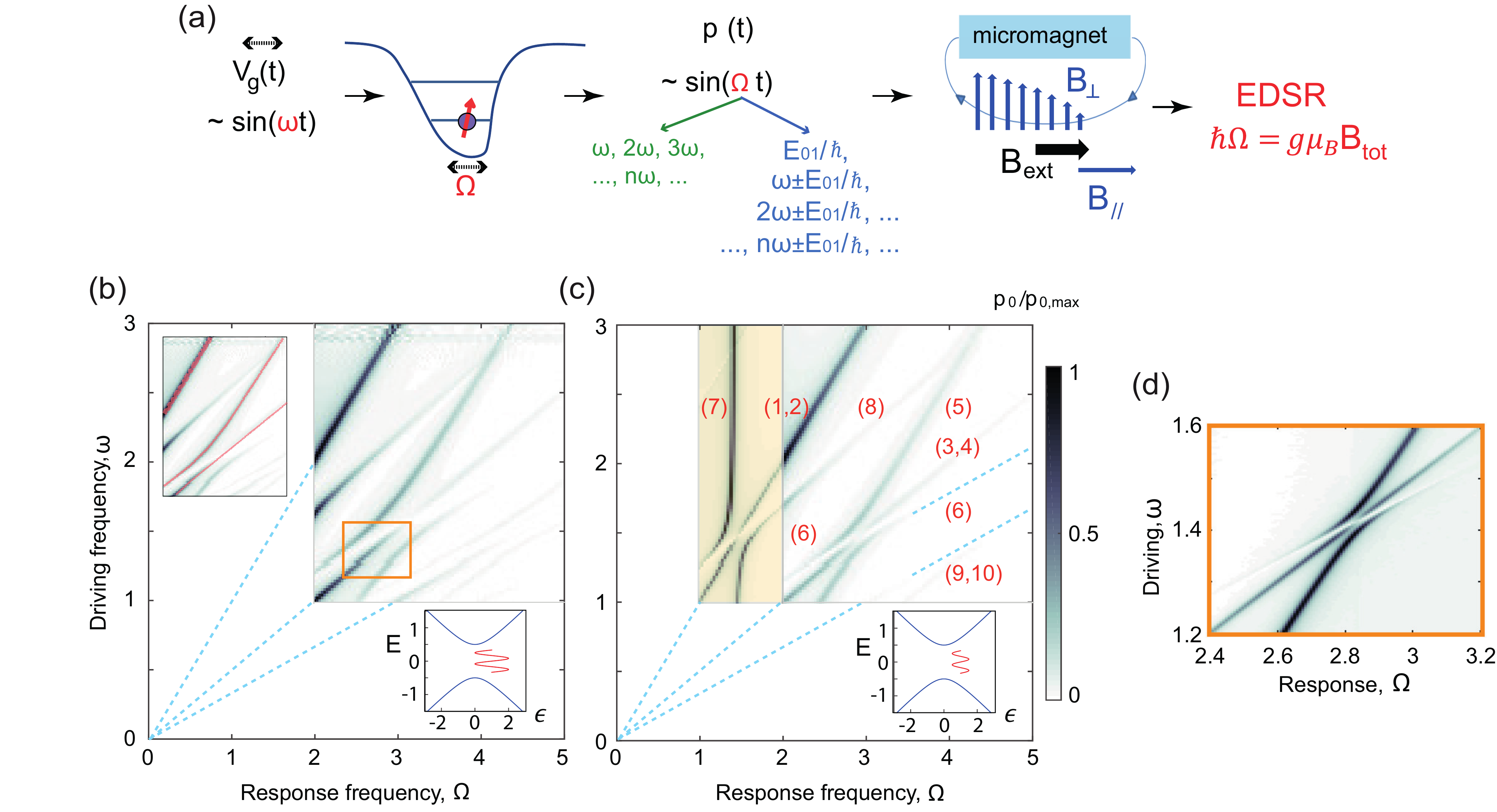}
\caption{Theoretical calculations of resonances in an ac-driven qubit with a low-lying orbital state induced by adiabatic and non-adiabatic processes. (a) Schematic illustrating that an excitation at driving frequency $\omega$ leads to several adiabatic (non-adiabatic) response frequencies of the electron dipole moment, $p(\Omega)=FFT[p(t)]$, listed in green (blue). The time-dependent electron dipole moment in turn produces spin flips due to the magnetic field gradient when $\hbar \Omega$ matches the Zeeman splitting.
(b) Simulated resonance spectrum of the ac-driven qubit model of Eqs.~(\ref{eq:realH}) and (\ref{eq:realepst}), setting $\hbar=1$. As described in the main text, the dynamics of the dipole moment $p(t)$, defined in Eq.~(\ref{eq:p0}), are solved in the time domain for the driving frequency $\omega$, then Fourier transformed to obtain the response frequency $\Omega$. Here we use $\epsilon_0=\epsilon_1=\Delta=1$.
The dashed lines indicate the positions of the fundamental resonance and its first two harmonics (top to bottom).
The lower inset shows the relation between the energy levels and the driving term.
The upper inset shows the same results as the main panel, with the experimentally relevant resonances highlighted (compare to Fig.~\ref{fig:Fig1b}).
(c) Resonance spectrum corresponding to the parameters $\epsilon_0=\Delta=1$ and $\epsilon_1=0.5$.
Here, the shaded region was normalized separately from the rest of the figure.  
The resonance features labeled 1-10 are discussed in the main text (see also Secs.~S.I.C and S.II.B of Supplemental Material).
(d) A blowup of the region shown in the center box of panel (b), using the parameters $\Delta=\epsilon_0=1$ and $\epsilon_1=0.11$, which gives good agreement with the level repulsion observed in the experiments shown in Figs.~\ref{fig:Fig1b}(a) and \ref{fig:Fig1b}(b).}
\label{fig:theory2} 
\end{figure*}

\begin{figure*}[ht!]
\includegraphics[width=\linewidth]{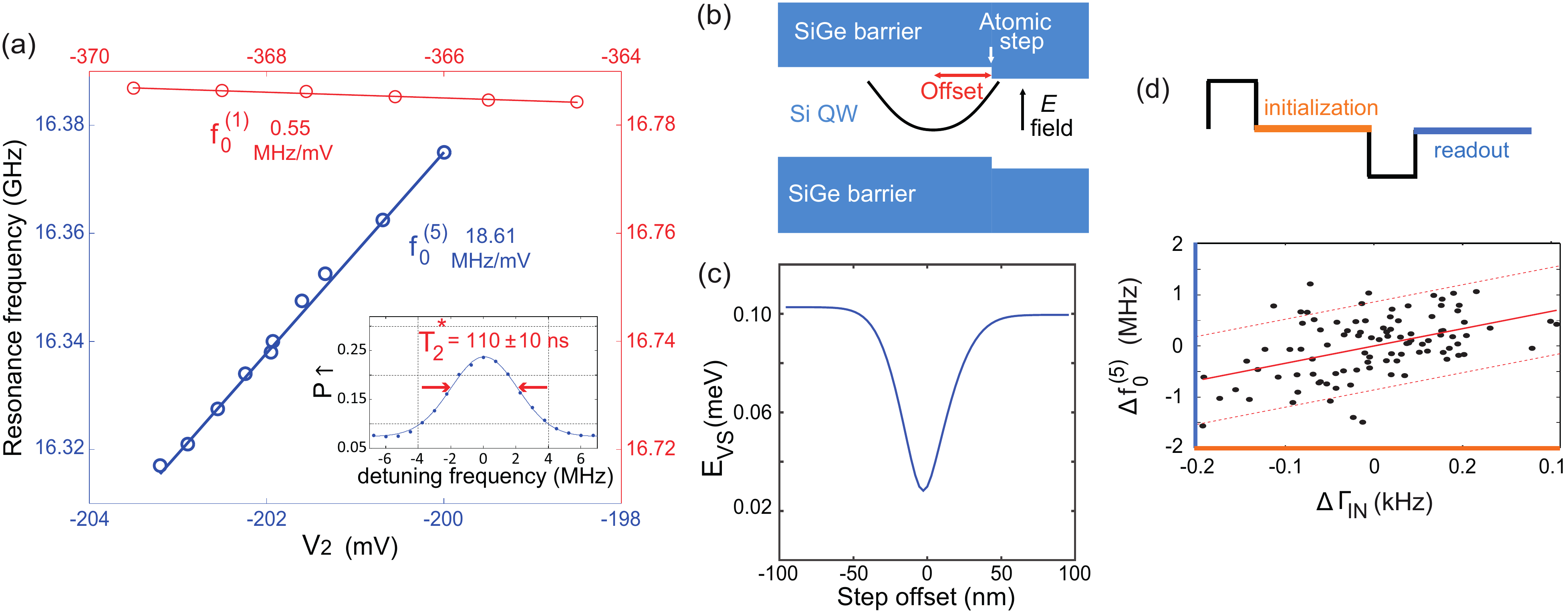}
\caption{Sensitivity of intra-valley and inter-valley spin transitions to electric fields. (a) Measured resonance frequency for the inter-valley ($f_0^{(5)}$ at $B^y_{ext}=$792 mT, blue data and axes) and intra-valley ($f_0^{(1)}$ at $B^y_{ext}=$550 mT, red data and axes) spin transition, as a function of gate voltage $V_2$. Inset: low-power continuous wave response of the inter-valley spin transition, with $T_2^\ast$ estimated from the line width (see also Fig.~S4). (b) Schematic representation of an atomic step in a Si/SiGe quantum well and of a quantum dot parabolic confinement potential (not to scale) laterally offset from the step. (c) Valley-orbit energy splitting found by a 2D tight binding calculation using the geometry shown in panel (b), a 13 nm wide quantum well barrier of 160 meV (corresponding to 30$\%$ Ge), a parabolic confinement potential for the dot of size $\sqrt{\langle x^2 \rangle } = 21.1$ nm (corresponding to an orbital energy splitting of $\hbar \omega = 0.45$ meV), and an electric field of 1.5$\times 10^6$ V/m (the experimental electric field is not well known). In the plots, a positive step offset corresponds to a step on the right-hand side of the dot. (d) Scatter plot showing two quantities measured every 400 s. The $x$ axis shows the average rate, $\Gamma_\text{IN}$, with which an electron tunnels into the quantum dot during qubit initialization (in orange in the inset); on the $y$ axis is the frequency of resonance (5) relative to its value averaged over the entire measurement (see also Fig.~S6). The continuous line represents a linear fit to the data through the point (0,0). The dashed lines represent the 95$\%$ confidence interval.
The distribution of the points in the scatter plot indicates that the two quantities are correlated. $B^x_\text{ext}=590$ mT, $B^y_\text{ext}=598.2$ mT, $\Delta f_0^{(5)}= f_0^{(5)}-15.6894$ GHz, $\Delta \Gamma_\text{IN} =\Gamma_\text{IN}-1.0762$ kHz.}
\label{fig:Fig5}
\end{figure*}

Our analysis builds upon the theory proposed by Rashba~\cite{Rashba2011}.
When a spin qubit is driven at a frequency $\omega$, it responds at one or more frequencies $\Omega$, which may be the same as $\omega$, but may also be different [see Fig.~\ref{fig:theory2}(a)]. 
Spin resonance is observed if (i) the spin is flipped \footnote{For resonance (5), the valley state must flip too; 
therefore, a spin-valley coupling mechanism is required \cite{Yang2013,Hao2014,Huang2014}.}, 
and (ii) $\hbar\Omega=E_Z$, where $E_Z=g\mu_BB_\text{tot}$ is the Zeeman splitting, $g$ is the Land\'{e} $g$-factor in silicon, $\mu_B$ is the Bohr magneton, and $B_\text{tot}$ is the total magnetic field.
In electric dipole spin resonance (EDSR), the spin flip requires a physical mechanism for the electric field to couple to the spin, such as spin-orbit coupling \cite{Tokura2006}.
In our experiment, an effective spin-orbit coupling due to the strong magnetic field gradient from the micromagnet is the mechanism responsible for spin flips \cite{Kawakami2014}.
Hence, we can say that EDSR and its associated spin dynamics provide a tool for observing the mapping $\omega\rightarrow\Omega$.
However, as discussed in \cite{Rashba2011}, EDSR does not determine the mapping; determining the resonant frequencies $\Omega$ requires including the essential non-linearity in the system, which in this case resides in the orbital sector of the qubit Hamiltonian. 
We therefore focus on the dynamics of the orbital sector of the Hamiltonian; the mechanism for spin flips is included perturbatively after the charge dynamics have been characterized.

The exact orbital Hamiltonian is difficult to write down from first principles, since it likely involves both orbital and valley components \cite{Kawakami2014}, and depends on the atomistic details of the quantum well interface \cite{Friesen2006,Friesen2010,Goswami2007}. Nonetheless, the features of the resonances in Fig.~\ref{fig:Fig1b} emerge quite naturally using a model with one low-lying orbital excited state. Referring to Fig.~\ref{fig:Fig1b}(c), in this model, the Hamiltonian for the orbital sector is described by a simple two-state Hamiltonian, which we write as
\begin{equation}
H=\frac{1}{2}(\epsilon\sigma_z+\Delta\sigma_x) . \label{eq:realH}
\end{equation}
Here, $\epsilon$ is a detuning parameter, $\Delta$ is the tunnel coupling between the generic basis states labeled $\ket{L}$ and $\ket{R}$, and $\sigma_{x}$ and $\sigma_z$ are Pauli matrices.
We consider a classical ac drive, applied to the detuning parameter:
\begin{equation}
\epsilon(t)=\epsilon_0+\epsilon_1\sin(\omega t) . \label{eq:realepst}
\end{equation}

If the quantum dot confinement were purely parabolic, then changing the detuning would not affect the energy splitting between the eigenstates. However, any nonparabolicity in the dot, which is unavoidable in real devices, will cause the energy splitting to depend on the detuning and will yield a nonlinear response to the driving term, Eq.~(\ref{eq:realepst}). In our Hamiltonian, this effect enters via the tunnel coupling $\Delta$, which causes the qubit frequency to depend on $\epsilon(t)$.

Our goal is to determine the response of the two-level system to this $\epsilon(t)$. We solve the time-dependent Schr\"odinger equation with the initial state $\psi(t=0)= \ket{0}$ representing the adiabatic ground state of Eqs.~(1) and (2) when $t = 0$, corresponding to $\epsilon_1 = 0$.
We assume that the basis states are coupled by the applied electric field because they have different spatial charge distributions, and study the time evolution of the instantaneous dipole moment of the ground state $\ket{0}$, defined as

\begin{equation}
p(t)=\frac{eL}{2} 
\left [ 
\left | \braket{\psi(t)}{L} \right |^2 - \left | \braket{\psi(t)}{R} \right |^2\ \right]. 
\label{eq:p0}
\end{equation}
Here, $L$ is the distance between the charge in states $\ket{L}$ and $\ket{R}$~\footnote{For a charge qubit, $L$ is the lateral separation between the two sides of the double quantum dot. For an orbital qubit, $L$ is the lateral separation of the center of mass of the two orbital states. For a pure valley qubit, $L$ is the vertical separation of the even and odd states ($\sim$0.16 nm). For a complicated system like a valley-orbit qubit with interface disorder, $L$ will have lateral and vertical components, with the lateral component being usually much larger than the vertical one. In this last case the exact length $L$ will depend on the specifics of the interface disorder; a reasonable guess would be $L\sim$0.5-5 nm (see, e.g. Ref. [28]).}.

Rashba has studied Hamiltonian (\ref{eq:realH}) perturbatively in the regime of weak driving
and high excitation frequency \cite{Rashba2011}. 
In Secs.~S.I and S.II of the Supplemental Material, we present a detailed exposition of our extensions of
these investigations into the strong driving regime relevant to resonances (5) and (6). We find that driving this transition involves non-adiabatic processes \cite{Landau1932,Zener1932,Stuckelberg1932} whereby the orbital state gets excited, in contrast to the subharmonics reported in \cite{Scarlino2015}, which as we show here involve only adiabatic processes in the charge sector.

Here, we present the results of numerical simulations in this regime and show that the results are consistent with the main features observed experimentally.
The dynamical simulations are performed by setting $\hbar=1$ and solving the Schr\"{o}dinger equation $i\partial \ket{\psi}/\partial t=H\ket{\psi}$ for Eqs.~(\ref{eq:realH}) and (\ref{eq:realepst}) and computing $p(t)$ as defined in Eq.~(\ref{eq:p0}) for a fixed driving frequency $\omega$~\footnote{For the Hamiltonian parameters indicated in the caption of Fig.~\ref{fig:theory2}(b), we use time steps in the range 0.061-0.073.}.
The resulting $p(t)$ is Fourier transformed, yielding a $p(\Omega)$ whose peaks reflect the resonant response.
Finally, we smooth $p(\Omega)$ by convolving it with a Gaussian of width 0.025 (in order to take into account noise, which is averaged in the experiment).
Because of the spin-orbit coupling (the position-dependent transverse magnetic field from the micro magnet), peaks in $p(\Omega)$ correspond to frequencies at which an ac magnetic field is generated that is resonant with the Zeeman frequency, $E_Z/\hbar=\Omega$, so spin flips will occur. The experiment measures the probability of a spin flip as a function of magnetic field; via the EDSR mechanism, resonances in this probability therefore occur when the peak locations in $p(\Omega)$ satisfy $g\mu_BB_\text{tot}=\hbar \Omega$.

\begin{figure*}[ht!]
\centering
\includegraphics[width=\linewidth] {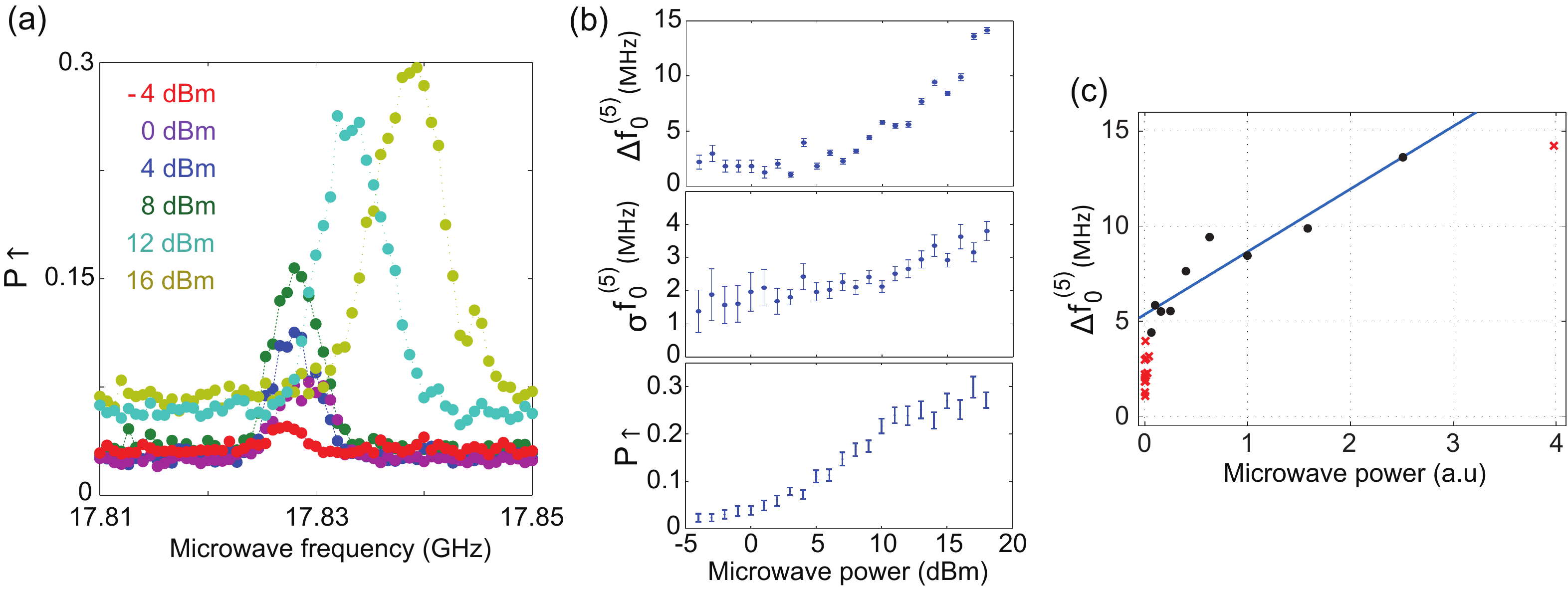}
\caption{
(a) Measured spin-up probability as a function of microwave power at $B^y_{ext}=846$ mT for a fixed microwave burst time of 50 $\mu$s, near the resonance condition for inter-valley spin-flip transition (5). For increasing power, the line does not only become taller and wider but also moves towards higher frequencies. Panel (b) summarizes from top to bottom the center frequency, width and height of the response. $\Delta f_0^{(5)}=f_0^{(5)}-17.825$ GHz. (c) Top panel of (b) replotted using a linear power scale ($x$ axis $\propto 10^{\mathbf{MW\,power(dBm)}/5}$). The blue line represents a linear fit to the black data points to the relation $\Delta E_\text{asymp}\propto \epsilon_1^2$ [Eq.~(S42)]. The points indicated by the red crosses are excluded from the fit.}
\label{fig:Fig7}
\end{figure*}

Figures~\ref{fig:theory2}(b) and (c) show the results of our simulations for $p(\Omega)$ as a function of both driving frequency $\omega$ and response frequency $\Omega$ over a range of parameters analogous to those shown in Fig.~\ref{fig:Fig1b}.
We discuss these transitions one by one (see also Sec.~S.I.C of the Supplemental Material). When driving at $\omega$ leads to a response in the dipole $p(t)$ at a frequency $\Omega$ such that $\omega=\Omega=E_Z$, a pure spin transition is induced. This corresponds to the fundamental resonances (1),(2). The first subharmonic response occurs when $\omega =\Omega/2 = E_Z/2$, which is the case of resonances (3),(4). For the second subharmonic, resonances (9),(10), we have $\omega = \Omega/3 = E_Z/3$. All the transitions discussed so far are adiabatic in the orbital sector. However, transitions that are non-adiabatic in the orbital sector are also allowed, i.e. transitions where an orbital excitation is involved (in the experiment, this would correspond to a valley excitation, the valley being the lowest energy orbital-like excitation). First, $\Omega = E_{01}$, a pure orbital excitation is induced (a pure valley transition with excitation energy $E_\text{VS}$ in the experiment), which is magnetic field independent and occurs over a wide range of $\omega$. This is transition (7) in Figs.~\ref{fig:theory2}(c). Next, resonance (5) runs parallel to the fundamental resonance, with $\omega = \Omega - E_{01}$ and furthermore $\Omega=E_Z$, so that $\omega = E_Z - E_{01}$. Thus driving at frequency $\omega$ induces a spin excitation and an orbital (valley) de-excitation, see the double red arrow in
Fig.~\ref{fig:Fig1b}(d). We call this process inter-valley spin resonance. Finally, resonance (6), which runs parallel to (9),(10), is characterised by $\omega = \Omega/3 + E_{01}/3$, again with $\Omega=E_Z$. Here, driving at $\omega$ leads to a spin excitation and an orbital excitation.
The upper inset of Fig.~\ref{fig:theory2}(b) highlights the particular resonances in the main figure that should be compared to the experimental data shown in Figs.~\ref{fig:Fig1b}(a) and ~\ref{fig:Fig1b}~(b).

An interesting feature of the resonances, observed both in the experiments and
theory [Figs.~\ref{fig:theory2}(b)-(d)], is the apparent `level repulsion' between resonance lines (5) and (6) that takes place near $B_\text{tot} = 480$ mT. This magnetic field value is much higher than $E_\text{VS}/g \mu_B \sim  B_\text{tot}=250$ mT, where the anti-crossing between the states $\left|v_1,\downarrow\rangle\right.$ and $\left|v_2,\uparrow\rangle\right.$ is expected to occur \cite{Yang2013} (see the red dashed line in Fig.~\ref{fig:Fig1b}(d)), but which is outside our measurement window, see above. Instead the observed `level repulsion' has a purely dynamical origin, as demonstrated by the fact that the anti-crossing is suppressed in Fig.~\ref{fig:theory2}(c), where the simulation parameters are identical to Fig.~\ref{fig:theory2}(b), except for a smaller driving amplitude $\epsilon_1$. 

In Sec.~S.II of the Supplemental Material, we develop a dressed-state theory to describe these strong-driving effects. In this formalism, the quasiclassical driving field of Eq.~(\ref{eq:realepst}) is replaced by a fully quantum description of the photon field and its coupling to the (valley)-orbital Hamiltonian of the quantum dot. The resulting dressed eigenstates describe the hybridized photon-orbital levels, and more generally, the hybridization of orbital, photon, and spin states. In this way, resonances (1) and (2) in Fig.~\ref{fig:theory2}(c) correspond to single-photon spin flips, while resonances (3) and (4) correspond to two-photon spin flips. Resonance (5) involves both a spin flip and a valley-orbit excitation and is parallel to resonances (1),(2), indicating that it is a single-photon process. Resonance (6) is parallel to (9),(10), indicating that it is a three-photon process. The physical mechanisms of the resonances are also indicated in Fig.~\ref{fig:Fig1b}(d). 

In principle, coupling occurs between all of the dressed states due to the effective spin-orbit coupling in our EDSR experiment. In practice however, the orbital-Rabi frequency is two orders of magnitude larger than the spin-Rabi frequency, so mode hybridization is only observed between resonance (5) and (6), resulting in the level repulsion. The magnitude of this repulsion provides a convenient way to determine the orbital-Rabi frequency, which cannot be measured directly due to the fast dephasing of the excited valley-orbit state (see Sec.~III). In Sec.~S.II of the Supplemental Material, we estimate this Rabi frequency to be about 0.2 GHz.

\section{Coherence of the inter-valley spin transition}

We now examine the possible origin of the ten times larger line width of resonance (5) compared to that of the pure spin-flip resonances (1) and (2). Given the partial valley nature of transition (5) and the strong valley-orbit coupling that is typical of Si/SiGe quantum dots \cite{Yang2013,Hao2014,Gamble2013}, a plausible candidate decoherence mechanism for this transition is electric field noise.

In order to study the sensitivity to electric fields of the respective transitions, we show in Fig.~\ref{fig:Fig5}(a) the dependence of the frequency of resonances (1) and (5) on the voltage applied to one of the quantum dot gates, $V_2$. Clearly, resonance (5) exhibits a much greater sensitivity to gate voltage than resonance (1): $\sim$ 18.5 MHz/mV for $f_0^{(5)}$, versus $\sim$ 0.5 MHz/mV for $f_0^{(1)}$. 
We also notice that the two resonance shifts as a function of $V_2$ have opposite sign, which indicates that different mechanisms are responsible. For resonance (1), we believe that the dominant effect of the electric field is the displacement of the electron wave function in the magnetic field gradient from the micromagnets \cite{Kawakami2014} (see Sec.~S.III.B of the Supplemental Material). This effect also contributes to the frequency shift of resonance (5), but presumably it is masked by the change in valley-orbit splitting ($E_\text{VS}$) resulting from the displacement of the electron wave function in the presence of interface disorder \cite{Shi2012,Friesen2010}. For instance, moving the electron towards or away from a simple atomic step at the Si/SiGe interface leads to a change of the valley-orbit energy splitting, as shown by the results of numerical simulations reported in Figs.~\ref{fig:Fig5}(b) and \ref{fig:Fig5}(c) \cite{Boross2016}. As expected, the simulations predict a minimum in the valley-orbit splitting when the wave function is centered around the atomic step, but interestingly it does not vanish, i.e.~the opposite signs for the valley-orbit splitting left and right of the atomic step do not lead to complete cancellation (see Sec.~S.IV of the Supplemental Material for a more detailed description).

The $\sim$35 times greater sensitivity of the spin-valley transition frequency to electric fields may contribute to its ten times larger line width compared to the intra-valley spin transition. The line width of the intra-valley spin transition is believed to be dominated by the 4.7$\%$ $^{29}$Si nuclear spins in the host material \cite{Kawakami2014}. The nuclear field also affects the spin-valley transition, but obviously only accounts for a small part of the line width here. We propose that the dominant contribution to the line width of resonance (5) is low-frequency charge noise.

Although not definitive, some evidence for this interpretation is found in Fig.~\ref{fig:Fig5}(d), which shows a scatter plot of $f_0^{(5)}$ and one of the dot-reservoir tunnel rates, simultaneously recorded over many hours (see Sec.~S.III.C of Supplemental Material for a more detailed description of the measurement scheme). The dot-reservoir tunnel rate serves as a sensitive probe of local electric fields, including those produced by charges that randomly hop around in the vicinity of the quantum dot (see Fig.~S6) \cite{Vandersypen2004}. The plot shows a modest correlation between the measured tunnel rate and $f_0^{(5)}$, suggesting that the shifts in time of both quantities may have a common origin, presumably low-frequency charge noise.

In this case, we can also place an upper bound on $T_2^*$ permitted by charge noise for the intra-valley spin transitions (1),(2) in the present sample (given the specific magnetic field gradient reported in \cite{Kawakami2014}). Indeed, due to the micro magnet induced gradient in the local magnetic field parallel to $B_\text{ext}$, the pure spin transitions are also sensitive to charge noise. Given that $T_2^* \sim$ 110 ns for transition (5) and the ratio of $\sim$ 35 in sensitivity to electric fields, charge noise in combination with this magnetic field gradient would limit $T_2^*$ to $\sim 3.8$ $\mu$s for transitions (1),(2). It is important to note that this is not an intrinsic limitation, as the stray field of the micro magnet at the dot location can be engineered to have zero gradient of the longitudinal component, so that to first order charge noise does not affect the frequency and $T_2^*$ of transitions (1),(2). At the same time, a strong gradient of the transverse component can be maintained, as is necessary for driving spin transitions.

Besides its strong sensitivity to static electric fields, we report a surprising dependence of the frequency of resonance (5) on microwave driving power [Figs.~\ref{fig:Fig7}(b) and \ref{fig:Fig7}(c)]. Increasing the driving power, the resonance not only broadens but also shifts in frequency, as in an a.c.~Stark shift \cite{Brune1994}. This dynamical evolution is very different from the case of the intra-valley spin resonance, which is power broadened but stays at fixed frequency \cite{Kawakami2014,Scarlino2015}. This frequency shift is, at least for a limited microwave power range, in line with the dynamical level repulsion captured by Eq.~(S42), where $\Delta E_\text{asymp}\propto \epsilon_1^2$ expresses the energy splitting between resonance (5) and its asymptote, $\hbar\omega=E_Z-E_{01}$. This relation is verified in Fig.~\ref{fig:Fig7}(c) for microwave powers of $9-17$ dBm.

Finally, we attempt to drive coherent oscillations using resonance (5) at high applied microwave power, recording the spin excited state probability as a function of the microwave burst time. Oscillations are not visible, indicating that the highest Rabi frequency we can obtain for resonance (5) is well below the corresponding $1/T_2^\ast$ of 110 ns. This is consistent with our estimate that the Rabi frequency is of the order of 10 kHz, based on the magnitude of the dynamical level repulsion seen in Figs.~\ref{fig:Fig1b}(a) and \ref{fig:Fig1b}(b) and the derivation in the Sec.~S.II.B of Supplemental Material.

\section{Conclusions}
Despite its simplicity, the electrical driving of a single electron confined in a single quantum dot can produce a complex spin resonance energy spectrum. This particularly applies for quantum dots realized in silicon, where the presence of the excited valley-orbit state, close in energy and strongly coupled to the ground state, introduces a substantial non-linearity in the system response to microwave electric fields. This allows us to observe a transition whereby both the spin and the valley state are flipped at the same time. We demonstrate how both static external electric fields and electrical noise influence the frequency of this inter-valley spin transition, dominating its coherence properties.

Much of the dynamics of the spin and valley transitions can be captured in a semi-classical picture, including driving using higher harmonics exploiting non-linearities. However, under intermediate or strong driving, new phenomena emerge that cannot be easily explained except in terms of dressed states that fundamentally involve a quantum mechanical coupling between photons and orbital or spin states. Here, we have provided experimental and numerical evidence for the existence of such dressed states of photons and valley-orbit states at strong driving. We have further estimated the strength of this valley-orbit to photon coupling by comparing our analytical theory to the experiments.

This work provides important experimental and theoretical insight in the role of inter-valley transitions for controlling spin dynamics in silicon based quantum dots. It also highlights the limitations of valley-based qubits in the presence of strong valley-orbit coupling, due to their sensitivity to electrical noise.

\section{Acknowledgments}
We acknowledge R. Schouten and M.~J. Tiggelman for technical support and the members of the Delft spin qubit team for useful discussions. Research was supported by the Army Research Office (W911NF-12-0607), a European Research Council Synergy grant, and the Dutch Foundation for Fundamental Research on Matter. E.K. was supported by a fellowship from the Nakajima Foundation. The development and maintenance of the growth facilities used for fabricating samples is supported by DOE (DE-FG02-03ER46028), and this research utilized NSF-supported shared facilities at the University of Wisconsin-Madison.

\beginsupplement

\section{Supplementary Materials}

This first section of these supplemental materials presents a theoretical discussion of the model used to understand
the electric dipole spin resonance experiments presented in the main text.
We show that experimental features are generic features of a driven four-level system, comprised of two valley-orbit and two spin degrees of freedom, with a tunnel coupling between the orbital states.
The model exhibits conventional resonances, including both fundamental and higher harmonics, as well as novel resonances involving photonically dressed orbital states.
We develop an analytical model that describes the key features of the hybridized dynamical states, based on a simple, 4D dressed-state Hamiltonian, and we use this model to determine the Rabi frequency for orbital excitations by fitting to the experimental data.

The main text reports EDSR measurements on an ac-driven spin qubit in a silicon quantum dot.
The system is driven in a microwave regime that is intermediate between weak and strong driving, allowing us to observe semiclassical phenomena, such as conventional spin resonances and spin-flip photon assisted tunneling (PAT)~\cite{Schreiber2011,Braakman2014}, which have been observed previously in quantum dots.
However, the experiments also probe new and intriguing phenomena that are purely quantum mechanical, involving the hybridization of photons, orbitals and spin.
Such effects are most conveniently described as dressed states.
By developing a formalism that is appropriate for our system, we can describe the photonically dressed states as pseudospins, and explain the hybridization effects by means of simple 2D and 4D Hamiltonians.
Moveover, although coherent oscillations of the photon-orbital-spin pseudospins are not observed in our experiments (in contrast with photon-spin pseudospins, whose coherent oscillations were reported in Ref.~\cite{Kawakami2014}), the dressed-state theory still allows us to extract the PAT Rabi frequency.

Our analysis is broken into two parts.
In Sec.~S.I, we describe a simple model for EDSR, and provide an intuitive discussion and overview of the resonances observed in the experiments. 
In Sec.~S.II, we develop a more technical, dressed-state formalism to describe the photonic dressing of the orbital and spin states in our system, which we use to explain and fit the experimental data. In Sec.~S.III we provide additional information on the experimental setup and on the measurements. In Sec.~S.IV we provide a more exhaustive description of the tight binding simulation realized in order to capture the evolution of the valley-orbit energy splitting as a function of the electron position in the SiGe/Si/SiGe quantum well [reported in Fig.~3(c)].

%
\section{S.I. Model of the Dynamics}\label{sec:Model}

When a spin qubit is driven at a frequency $\omega$, it responds at one or more frequencies $\Omega$, which may be the same as $\omega$, but may also be different. 
Spin resonance is observed if (i) the spin is flipped, and (ii) $\hbar\Omega=E_Z$, where $E_Z=g_B\mu_BB$ is the Zeeman splitting, $g_B$ is the Land\'{e} $g$-factor in silicon, $\mu_B$ is the Bohr magneton, and $B$ is the applied magnetic field.
The spin flip requires a physical mechanism, such as spin-orbit coupling in EDSR~\cite{Tokura2006}.
Indeed, the strong magnetic field gradient of the micromagnet introduces such
a coupling~\cite{Tokura2006,Kawakami2014}.
Rashba has suggested that the mapping $\omega\rightarrow\Omega$ occurs entirely within the orbital sector, and that EDSR simply provides a tool for observing this mapping~\cite{Rashba2011}.
In this Supplemental Section, we adopt Rashba's orbital-based model as our starting point.

Several theoretical explanations have been put forth to explain strong-driving phenomena in EDSR, such as the generation of higher harmonics~\cite{Rashba2011,Stehlik2014,Szechenyi2014,Danon2014}, and other, more exotic effects such as an even/odd harmonic structure~\cite{Stehlik2014}.
For silicon dots, the exact orbital Hamiltonian is difficult to write down from first principles, since it involves both orbital and valley components~\cite{Kawakami2014}, and depends on the atomistic details of the quantum well interface~\cite{Friesen2006,Goswami2007,Friesen2010}.
Nonetheless, the dynamics can be understood using a minimal model for investigating the physics of the orbital sector, given by the two-level system 
\begin{equation}
H=\frac{1}{2}(\epsilon\sigma_z+\Delta\sigma_x) , \label{eq:H}
\end{equation}
where $\sigma_{x}$ and $\sigma_z$ are Pauli matrices.
The basis states in this model may correspond to excited orbitals in a single quantum dot~\cite{Tokura2006}, valley states in a single dot~\cite{Friesen2006}, localized orbital states in a double dot~\cite{Rashba2011}, or a hybridized combination of these systems~\cite{Friesen2010}.
In each case, the two basis states have distinct charge distributions that can support EDSR through the mechanism described by Tokura et al.~\cite{Tokura2006}

Equation~(\ref{eq:H}) is, in fact, the most general theoretical model of a two-level system.
Mapping the theoretical model onto a real, experimental system requires specifying the dependence of the effective detuning parameter $\epsilon$ and the effective tunnel coupling $\Delta$ on the voltages applied to the top gates for a given device.
To take a simple example that is relevant for silicon quantum dots, we consider such a mapping for the basis comprised of valley states.
In this case, $\epsilon$ represents a valley splitting, which can be tuned with a vertical electric field, while $\Delta$ represents a valley coupling, which can depend on the interfacial roughness of the quantum well and other materials parameters~\cite{Gamble2013}.
The centers of mass of the charge distributions of the basis states in this example differ by about a lattice spacing in the vertical direction.
It is likely that the silicon dots in the experiments reported here experience an additional valley-orbit coupling, which induces a lateral variation of the charge distributions.
In this case, the gate voltage dependence of $\epsilon$ and $\Delta$ is quite nontrivial; however Eq.~(\ref{eq:H}) should still provide a useful theoretical description.
For convenience in the following analysis, we adopt the picture of a charge qubit, for which the basis states correspond to ``left" ($\ket{L}$) and ``right" ($\ket{R}$).
The eigenstates of $H$ are then given by
\begin{gather}
\ket{0}=\frac{\Delta}{\sqrt{\Delta^2+(\epsilon+E_{01})^2}}\ket{L}
-\frac{\epsilon+E_{01}}{\sqrt{\Delta^2+(\epsilon+E_{01})^2}}\ket{R} , \label{eq:0}  \\
\ket{1}=\frac{\Delta}{\sqrt{\Delta^2+(\epsilon-E_{01})^2}}\ket{L}
-\frac{\epsilon-E_{01}}{\sqrt{\Delta^2+(\epsilon-E_{01})^2}}\ket{R} , \label{eq:1}
\end{gather}
and the energy splitting is given by
$E_{01}=\sqrt{\Delta^2+\epsilon^2}$.

To begin, we consider a classical ac driving field, applied to the detuning parameter:
\begin{equation}
\epsilon(t)=\epsilon_0+\epsilon_1\sin(\omega t) . \label{eq:epst}
\end{equation}
Our goal is to determine the response of the two-level system to this $\epsilon(t)$.
In contrast with Rashba, we will not limit our analysis to the perturbative regime in
which $\epsilon_1,E_{01}\ll \hbar\omega$, since the experiments reported here are not in that regime.
To proceed, we note that the basis states couple differently to the applied electric field because they have different spatial charge distributions.
We specifically consider the time evolution of the dipole moment of the ground state $\ket{0}$, defined as
\begin{equation}
p=\frac{eL}{2}\bra{0}\sigma_z\ket{0}=\frac{eL}{2} \left[ \left|\braket{0}{L}\right|^2 - \left|\braket{0}{R}\right|^2 \right]  , \label{eq:p0}
\end{equation}
where $L$ is the distance between the center of mass of the charge in states $\ket{L}$ and $\ket{R}$.
Throughout the discussion that follows, when connecting the output of the
model to the experimental data, we will identify $E_{01}$ with the valley-orbit splitting, $E_\text{VS}$.

\subsection{A. Adiabatic Effects} \label{sec:adiabatic}
Two fundamentally different types of response to the driving are observed in $p(t)$:  adiabatic vs.\ nonadiabatic.
We first consider adiabatic processes.
The adiabatic eigenstates of the Hamiltonian, given in Eqs.~(\ref{eq:0}) and (\ref{eq:1}) yield
a corresponding dipole moment, computed from Eq.~(\ref{eq:p0}), of
\begin{equation}
p=\frac{eL}{2} \left[ \frac{\Delta^2-(\epsilon+E_{01})^2}{\Delta^2+(\epsilon+E_{01})^2} \right] 
=-\frac{eL\epsilon}{2E_{01}} .
\label{eq:p0adiabatic}
\end{equation}
Substituting Eq.~(\ref{eq:epst}) into Eq.~(\ref{eq:p0adiabatic}), we see that the adiabatic response of $p(t)$ is a nonlinear function of $\sin(\omega t)$.
For example, in the weak driving limit, we can expand Eq.~(\ref{eq:p0adiabatic}) to second order in the small parameter $\epsilon_1/E_{01}$, giving
\begin{eqnarray}
p(t)&\simeq& \text{(const.)}-\left( \frac{eL\Delta^2}{2\bar{E}_{01}^3} \right) \epsilon_1 \sin(\omega t) 
\label{eq:p0expand} \\ \nonumber && \hspace{.5in} 
-\left(\frac{3eL\Delta^2\epsilon_0}{8\bar{E}_{01}^5}\right) \epsilon_1^2\cos(2\omega t) +\dots .
\end{eqnarray}
Here, $\bar{E}_{01}$ denotes the energy splitting when $\epsilon=\epsilon_0$.
We see that the nonlinear dependence of the dipole moment on $\epsilon_1$ translates into a series of response terms at frequencies $\Omega=\omega, 2\omega, 3\omega, \dots$.
Since spin resonances only occur when $\hbar\Omega=E_Z$, these should be observed as subharmonics of the fundamental Zeeman frequency: $\omega = E_Z/\hbar, E_Z/2\hbar, E_Z/3\hbar,\dots$.
For the two-level system defined by Eq.~(\ref{eq:H}), we note that all resonances depend on the presence of a coupling between the basis states, $\Delta$.
In Fig.~1(d) of the main text, the fundamental resonance condition and its second harmonic are sketched as the green and blue vertical arrows respectively.
We note that adiabatic resonances can be observed in both the strong and weak driving regimes, although the higher harmonics may be suppressed for weak driving, as consistent with Eq.~(\ref{eq:p0expand}).

\subsection{B. Nonadiabatic Effects}\label{sec:nonadiabatic}
We next consider nonadiabatic, or Landau-Zener (LZ) processes~\cite{Landau1932,Zener1932,Stuckelberg1932}, which always involve an excitation (or deexcitation) of an orbital state. 
For the two-state Hamiltonian of Eq.~(\ref{eq:H}), LZ excitations can arise in two ways: (i) a sudden pulse of a control parameter (e.g., $\epsilon$), or (ii) strong periodic driving of the control parameter [e.g., Eq.~(\ref{eq:epst})].
In the first case, the qubit is suddenly projected onto a new adiabatic eigenbasis, which induces a Larmor response at the frequency $\Omega=\pm E_{01}/\hbar$. 
In the second case, fast periodic excursions of $\epsilon(t)$, which does not need to occur at the resonant frequency, can also produce a similar response.
Such `conventional' LZ excitation processes are indicated by the black vertical arrow labeled (7) in Fig.~1(d) of the main text~\cite{NoteS1}.

In the experiments reported here, simple LZ orbital excitations are undetectable, because the excitation energy $E_{01}$ is smaller than the Fermi level broadening.
However, the EDSR spin-flip mechanism allows us to generate and detect more complex processes at higher energies.
The resonance conditions indicated by the red and yellow vertical arrows in Fig.~1(d) are particularly important for these experiments; when combined with a spin flip, the LZ orbital excitation corresponds to resonances (5) and (6) shown in Figs.~1(a) and (b).
Resonances of this type occur when $\hbar\Omega =m\hbar\omega\pm E_{01}$, corresponding to the driving frequencies $\omega = (E_Z\pm E_{01})/\hbar , (E_Z\pm E_{01})/2\hbar , (E_Z\pm E_{01})/3\hbar ,\dots$.
Since these resonance lines are parallel to the conventional spin resonances, Rashba has called them ``satellites"~\cite{Rashba2011}, although he studied a different driving regime for which no satellites were observed in the experiments. 

There are two main theoretical approaches for describing such nonadiabatic phenomena.
Since the orbital excitation involves an LZ process, it must be described quantum mechanically.
However, the ac driving field can be described with a semiclassical, adiabatic theory like the one described in the previous section.
In this approach, the qubit gains a ``St\"uckelberg phase"~\cite{Stuckelberg1932} each time it passes through an energy level anticrossing, resulting in interference effects that can be described using standard LZS theory~\cite{Shevchenko2010,NoteS2}, or other methods based on the Floquet formalism~\cite{Shirley1965,Chu2004}.
The alternative theoretical approach, described below, uses a dressed state formalism that encompasses all of the same interference effects, and plays a key role in understanding the detailed behavior of resonances (5) and (6).
In this case, the process can be described fully quantum mechanically as an orbital excitation combined with a photon absorption.
In either description, the orbital excitation is followed by a spin-flip caused by the orbital dynamics.

\subsection{C. Simulation Results}\label{sec:simulations}
Results of numerical simulations of Eq.~(\ref{eq:H}) are presented in Figs.~2(b)-(d) of the main text.
Here, we discuss the physical interpretation of these results in more detail.

\begin{figure}[t]
\includegraphics[width=2.5in]{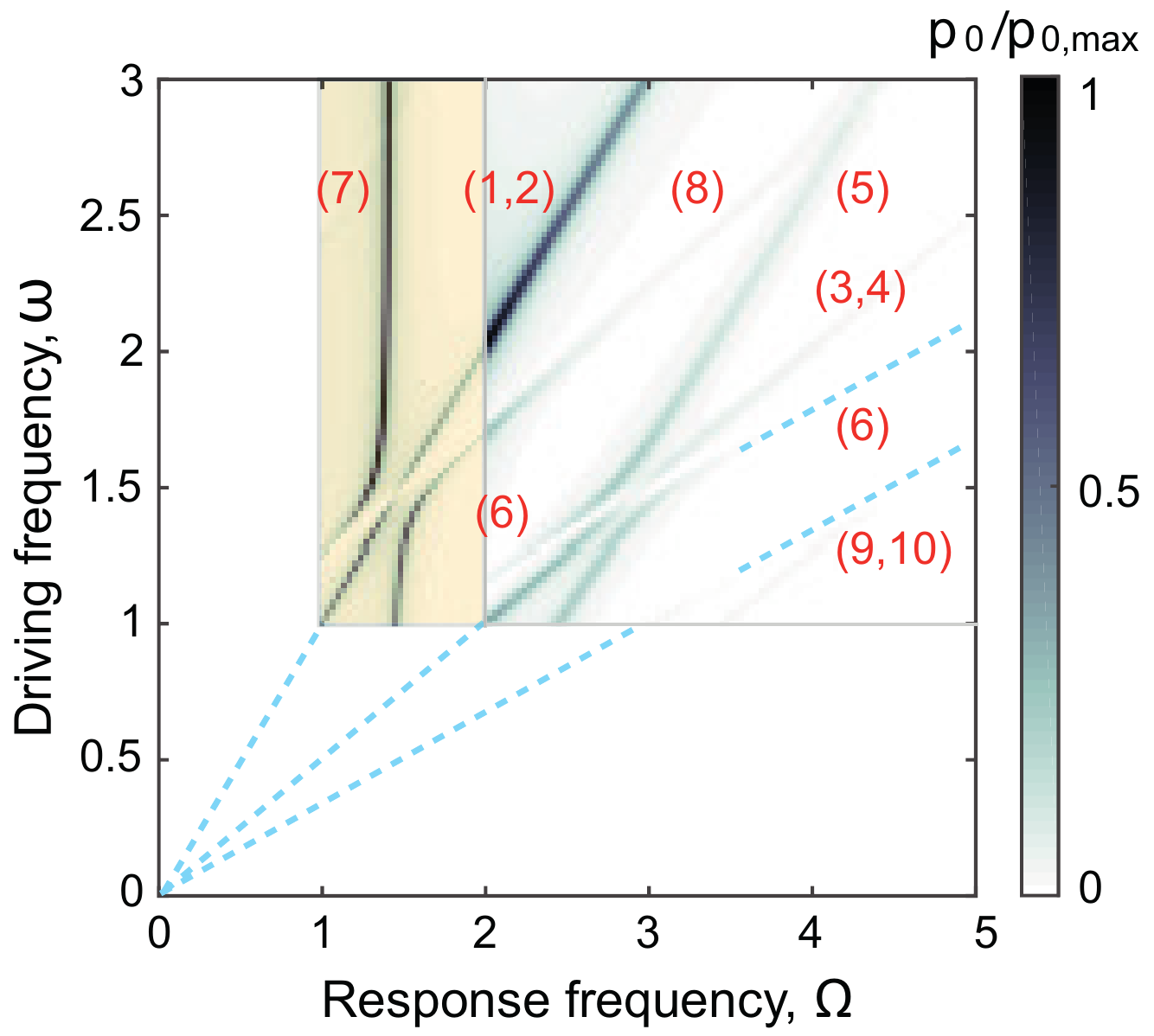}
\caption{
Simulation results, reproduced from Fig.~2(c) in the main text.
Here, the resonances labeled (1) through (10) are described in Sec.~S.I.C.
}
\label{fig:theorylabels} 
\end{figure}

Figure~2 in the main text shows the results of our simulations for $p(\Omega)$ as a function of both driving frequency $\omega$ and response frequency $\Omega$ over a range of parameters analogous to those in Fig.~1(a).
Figures~2(b)-(d) correspond to three different driving amplitudes, with panel (d) chosen to match the experimental results in Fig.~1(b).
The dashed blue lines correspond to the conventional ESR signals (the fundamental resonance, and the first two harmonics, top to bottom), which are all visible in the simulations.
In Fig.~2(c) (reproduced in Fig.~\ref{fig:theorylabels}), these are labeled (1),(2), (3),(4), and (9),(10).
The inset on the left-hand side of Fig.~2(b) highlights the particular resonances that should be compared to the experimental data in Fig.~1(a).

In addition to conventional ESR resonances, several LZ resonances can be seen in the simulations.
Three LZ resonances, labeled (5), (6) and (8) in Fig.~2(c), are observed between the fundamental resonance and the second harmonic.
At high frequencies, we note that resonance (8) has the same slope as the second harmonic (3),(4), while resonance (5) has the same slope as the fundamental (1,2).
More explicitly, these lines correspond to the resonance conditions given by $\omega = (E_Z+E_{01})/2\hbar$ and $\omega=(E_Z-E_{01})/\hbar$, respectively.
The latter resonance is clearly visible in the experiment, and corresponds to the resonance labeled (5) at high frequencies in Fig.~1. Resonance (6) has the same slope as the third harmonic (9),(10) and is visible in the experiment at the lowest frequencies.
Its frequency is given by $\omega = (E_Z+E_{01})/3\hbar$.

An interesting feature of the LZ resonances, which is also observed in the experiments [see Fig.~1(b)], is the apparent level repulsion between the resonance peaks, for instance between (5) and (6).
Several of these anticrossings are seen in Fig.~2.
The effect is purely dynamical, as demonstrated by the fact that the anticrossing is suppressed in going from panels (b) to (d), as consistent with the smaller driving amplitudes.
In the dressed-state formalism, these dynamical modes correspond to resonances of the hybridized photon-orbital states, where the state composition changes suddenly.

Several excitations are observed in the simulations of Fig.~2 that are not also observed in the experiments.
For example, the vertical line labeled (7) in Fig.~2(c) corresponds to the nonresonant (Larmor) LZ excitation discussed in Sec.~S.I.B.
This excitation occurs at an energy $\hbar\Omega$ that is too small to be detected experimentally.
Resonance (8) is suppressed in the simulations, compared to resonance (5), and it is not observed at all in the experiments.
A level anticrossing is observed between resonances (7) and (8) in the simulations, which also occurs outside the experimental measurement window.
Several other features can be observed in the lower-right corner of Fig.~2(b), which are very faint, and can also be classified according to the resonance scheme discussed above.

\section{S.II. Dressed States}\label{sec:Dressed}

Previous approaches to strong driving have assumed that the driving amplitude $\epsilon_1$ and the driving frequency $\hbar\omega$ are both much larger than the minimum energy gap $\Delta$~\cite{Oliver2005}.
However, the simulation results shown in Fig.~2, which provide a good descrption of the experiments in Fig.~1, suggest that our system falls into an intermediate regime, where the dimensionless parameters all have similar magnitudes, $\Delta,\epsilon_0,\hbar\omega\sim 1$, and $\epsilon_1\sim 0.1$. 
Below, we develop a new dressed state formalism that is appropriate for such an intermediate parameter regime, while treating $\epsilon_1$ as weakly perturbative.
This formalism successfully describes all the resonance features observed in Figs.~1 and 2.
In keeping with our emphasis on orbital physics, we will first consider only the charge and photon sectors, and introduce the spin afterwards.

\subsection{A. Orbital-Photon System}\label{sec:charge}

The time-dependent Hamiltonian given by Eqs.~(\ref{eq:H}) and (\ref{eq:epst}) is semiclassical (i.e., the electric field is treated classically). The same Hamiltonian was investigated in Ref.~\cite{Oliver2005}, in the context of strong driving.
A fully quantum version of the same problem was considered in Refs.~\cite{Nakamura2001} and \cite{Wilson2007}, also in the context of strong driving.
The solutions to the semiclassical and quantum Hamiltonians are identical, suggesting that the two approaches are interchangeable.
The quantum version is more elegant however, and it is more directly compatible with the dressed state formalism.
We therefore adopt the quantum Hamiltonian as our starting point, rather than Eq.~(\ref{eq:H}).
The full quantum Hamiltonian is given by
\begin{equation}
H=\frac{1}{2}(\epsilon_0\sigma_z+\Delta\sigma_x)+\hbar\omega a^\dagger a+g\sigma_z(a+a^\dagger) .
\label{eq:quantumH}
\end{equation}
Here, $a^\dagger$ and $a$ are the creation and annihilation operators for microwave photons of frequency $\omega$, and $g$ is the electron-photon coupling constant (we ignore the vacuum fluctuations of the electric field). The semiclassical and quantum Hamiltonians are related through the correspondence~\cite{Nakamura2001}
\begin{equation}
g=\epsilon_1/4\sqrt{\langle N\rangle} , \label{eq:gcouple}
\end{equation}
where $N=a^\dagger a$ is the photon number operator.
We expect that $\langle N\rangle\gg 1$ for gate-driven microwave fields.
The ac detuning amplitude can be related to the ac voltage applied to a top gate, $V_\text{ac}$, through the relation $\epsilon_1=eV_\text{ac}(C_g/C_\Sigma)$, where $C_g$ is the capacitance between the dot and the top gate and $C_\Sigma$ is the total capacitance of the dot \cite{Nakamura2001}.

It is important to note the form of the coupling term in Eq.~(\ref{eq:quantumH}), which arises because the ac signal is applied to the detuning parameter.
The resulting Hamiltonian cannot be solved exactly, and several approaches have been used to simplify the problem.
In the weak driving limit, it is common to combine a unitary transformation with a rotating wave approximation to reduce the coupling in Eq.~(\ref{eq:quantumH}) to a transverse form:
$g(\sigma_+a+\sigma_-a^\dagger)$~\cite{Childress2004}.
The latter has been used to describe the interactions between a dot and a microwave stripline~\cite{Frey2012}.
Alternatively, in the strong driving limit, it is common to retain the longitudinal coupling form of Eq.~(\ref{eq:quantumH}) while treating $\Delta$ as a perturbation.
This approach has been used to describe the dynamics of a dot in a gate-driven microwave field~\cite{Nakamura2001,Wilson2007}.
In the intermediate driving regime, which is most relevant for the experiments reported here, neither of these approaches is appropriate.

As discussed in Supplemental Sec.~S.I.B, the LZ satellite peaks in Fig.~2 are displaced vertically from the conventional spin resonances (the center peaks) by $E_{01}/m\hbar$~\cite{NoteS3}. For example, the satellite peak labeled (5) in Fig.~2(c) is displaced by $\sqrt{2}$ from the fundamental resonance (1),(2), while the satellite labeled (8) is displaced by $\sqrt{2}/2$ from the conventional second harmonic (3),(4).
This suggests that we should first diagonalize the uncoupled ($g=0$) Hamiltonian, giving $\tilde H_0=-\frac{E_{01}}{2}\tilde\sigma_z+\hbar\omega a^\dagger a$, where $\tilde\sigma_z$ is the Pauli matrix along the orbital quantization axis.
The eigenstates of $\tilde H_0$ are labeled $\ket{c}\ket{N}$, where $(\tilde\sigma_z/2)\ket{c}=c\ket{c}$ and $c=\pm 1/2$, and the energy eigenvalues are given by $-c E_{01}+N\hbar\omega$.
The full Hamiltonian becomes
\begin{equation}
\tilde H=-\frac{E_{01}}{2}\tilde\sigma_z+\hbar\omega a^\dagger a
+g(\cos\theta\,\tilde\sigma_x-\sin\theta\,\tilde\sigma_z)(a+a^\dagger) ,
\label{eq:H'}
\end{equation}
where we have defined $\tan\theta=\epsilon_0/\Delta$.

To dress the orbital states, we consider the Hamiltonian $\tilde H_l=\tilde H_0-g\sin\theta \tilde\sigma_z(a+a^\dagger)$, whose coupling is purely longitudinal.
This Hamiltonian has exact solutions, given by~\cite{TannoudjiBook},
\begin{equation}
\ket{c;N}_l=\exp [-2c g\sin\theta(a^\dagger -a)/\hbar\omega ]\ket{c}\ket{N} , \label{eq:dressed}
\end{equation}
with the corresponding energies
\begin{equation}
E_N^c=-c E_{01}  +N\hbar\omega -\frac{(g\sin\theta)^2}{\hbar\omega} . \label{eq:dressedE}
\end{equation}
Here, the subscript $l$ indicates a longitudinal dressed state.

Evaluating the full Hamiltonian $\tilde H$ in the dressed state basis, we obtain diagonal terms given by Eq.~(\ref{eq:dressedE}), and off-diagonal terms given by
\begin{gather}
_l\bra{c;N}g\cos\theta\,\tilde\sigma_x(a+a^\dagger) \ket{c;N+m}_l = 0  , \label{eq:ME1}  \\
_l\bra{-c;N}g\cos\theta\,\tilde\sigma_x(a+a^\dagger) \ket{c;N}_l = 0  , \label{eq:ME2}  \\
_l\bra{-c;N}g\cos\theta\,\tilde\sigma_x(a+a^\dagger) \ket{c;N+m}_l =  h\nu_{N,m}/2 \nonumber
\\ \hspace{2.1in} (\text{with}\,\, m\geq 1) \label{eq:ME3} ,
\end{gather}
where
\begin{multline}
\frac{h\nu_{N,m}}{2}=g\cos\theta e^{-2\alpha^2}\sum_{p=0}^{N}
\left[1+\frac{4\alpha^2(N-p)}{(m+p)(m+p+1)}\right] 
\\  \times
\frac{(-1)^{m+p-1}(2\alpha)^{m+2p-1}\sqrt{N!(N+m)!}}{p!(m+p-1)!(N-p)!} , \label{eq:nu}
\end{multline}
and
\begin{equation}
\alpha=-\frac{2cg\sin\theta}{\hbar\omega} . \label{eq:alpha}
\end{equation}
We note that Eq.~(\ref{eq:ME1}) follows from the facts that (i) the orbital character of $\ket{c;N}_l$ is the same as $\ket{c}\ket{N}$, and (ii) $\bra{c}\tilde\sigma_x\ket{c}=0$.
Equations~(\ref{eq:ME2}) and (\ref{eq:ME3}) are obtained using standard techniques~\cite{TannoudjiBook}.

We can obtain an approximate form for the Rabi frequency $\nu_{N,m}$ for orbital oscillations.
Since our simulations indicate that $\epsilon_1\ll \hbar\omega$, and we have assumed that $\langle N\rangle\gg 1$, Eqs.~(\ref{eq:gcouple}) and (\ref{eq:alpha}) imply that $\alpha\ll 1$.
Since the distribution of photon number states comprising $\ket{c;N}_l$ is sharply peaked around $\langle N\rangle$, we can approximate $N\simeq \langle N\rangle$, yielding
\begin{equation}
\frac{h\nu_{N,m}}{2}\simeq g\cos\theta\left[1+\frac{4\alpha^2\langle N\rangle}{m(m+1)}\right]
\frac{(-2\alpha)^{m-1}}{(m-1)!}\langle N\rangle^{m/2} . \label{eq:nulargeN}
\end{equation}
Finally, in the limit $\alpha\sqrt{\langle N\rangle}\ll1$, we find
\begin{equation}
\nu_{N,m}\simeq \frac{\omega/2\pi}{(m-1)!\tan\theta}\left(\frac{\epsilon_1\sin\theta}{2\hbar\omega}\right)^m .
\label{eq:nulimit}
\end{equation}
Thus, ignoring geometrical factors, we obtain the general scaling relation $\nu_{N,m}\simeq \epsilon_1^m/(\hbar\omega)^{m-1}$.

The dressed state formalism is convenient for identifying and analyzing resonant phenomena.
For example, the resonance condition for an $m$-photon orbital excitation is simply given by $E_{N+m}^{+1/2}\simeq E_N^{-1/2}$, or $m\hbar\omega\simeq E_{01}$.
When this condition is satisfied, the full Hamiltonian approximately decouples into a set of 2D manifolds given by
\begin{equation}
H_\text{2D}=\begin{pmatrix}
E_{01} & \frac{h\nu_{N,m}}{2} \\
\frac{h\nu_{N,m}}{2} & m\hbar \omega \end{pmatrix} , \label{eq:H2D}
\end{equation}
where we have subtracted a constant diagonal term.
For a system initially in the orbital state $c=1/2$, Eq.~(\ref{eq:H2D}) describes oscillations between $c=1/2$ and $-1/2$, with the Rabi frequency $\nu_{N,m}$.
These resonances are referred to as photon assisted tunneling (PAT), and are well known in quantum dots~\cite{Kouwenhoven1994}.
In the present context, ``tunneling" refers to the generalized tunneling parameter $\Delta$ in Eq.~(\ref{eq:H}), which represents the coupling between the different orbital states.

\subsection{B. Spin-Orbital-Photon System}\label{sec:spin}

Following Rashba \cite{Rashba2011}, we have argued that the experimental EDSR spectrum can be explained entirely within the orbital sector.
The argument is summarized as follows:  
(1) the orbital (two-level) system is driven at a frequency $\omega$; 
(2) nonlinearities in the orbital system generate responses at multiples of $\omega$, corresponding to multiphoton processes; 
(3) strong driving modifies the response by generating additional satellite peaks at frequencies $\pm E_{01}/\hbar$; 
(4) the response frequencies in the orbital sector are transferred directly to the spin sector, where they become the driving frequencies for spin rotations via the EDSR mechanism.
By associating the response frequency $\Omega$ in the orbital system with the characteristic frequency $g\mu_BB/\hbar$ in the spin system, we were able to obtain excellent agreement with the experiments, without invoking any other assumptions about the spin physics.
In an  alternative explanation of the spin harmonics, we could consider unrelated nonlinearities in the spin sector, independent of the orbital sector.
Two arguments against such an approach are as follows.
First, since nonlinearities occur in the orbital system anyway, Occam's razor suggests using the simplest, weak-driving model for the spin dynamics.
Second, while the spin Rabi frequencies obtained in our previous EDSR experiments are $<10$~MHz~\cite{Scarlino2015,Kawakami2014}, the Rabi frequencies for orbital oscillations are $>100$~MHz (see Sec.~S.II.C); it is therefore very reasonable to consider strong, photon-mediated driving for the orbital system, but weak driving for the spin system, mediated by the orbital dynamics.

We can formalize such a ``simple-spin" model by adopting a standard spin Hamiltonian,
\begin{equation}
H_\text{spin}=-\frac{1}{2} g_B\mu_B \hat{\mathbf B}(\hat{\mathbf r}[\sigma_z])\cdot \bm \tau .
\label{eq:Hspin}
\end{equation}
Here, $\hat{\mathbf B}$ is a superoperator representing the inhomogeneous magnetic field, $\hat{\mathbf r}$ is the electron position operator, which is related linearly to the dipole moment operator in Eq.~(\ref{eq:p0}),  and $\bm \tau$ is the Pauli operator for a (spin-1/2) electron.
The explicit dependence of $\hat{\mathbf r}$ on $\sigma_z$ indicates that $\hat{\mathbf r}$ acts only on the orbital variables.
We emphasize that there is no direct coupling between spins and photons in this model.

The quantization axis for the spins is defined as the average of the magnetic field operator:
\begin{equation}
{\mathbf B}_\text{avg}  = \langle \hat{\mathbf B}(\hat{\mathbf r}) \rangle ,
\end{equation}
with the corresponding Zeeman energy, $E_Z=g\mu_BB_\text{avg}$.
The deviation of $\hat{\mathbf B}(\hat{\mathbf r})$ from its average value (e.g., due to driving) is given by
\begin{equation}
\delta \hat{\mathbf B}(\hat{\mathbf r})= \hat{\mathbf B}(\hat{\mathbf r})-{\mathbf B}_\text{avg}  .
\end{equation}
Clearly, $\delta \hat{\mathbf B}(\hat{\mathbf r})$ contains information about the magnetic field gradient, which is the source of spin-orbit coupling in EDSR.
Transforming ${\bm\tau}$ to the quantization frame yields $\tilde\tau_z=\bm\tau \cdot \mathbf{B}_\text{avg}/B_\text{avg}$, where $(\tilde\tau_z/2)\ket{s}=s\ket{s}$ and $s=\pm 1/2$.
Note that we have identified $s=+1/2$ as the low-energy spin state, in analogy with the orbital system.

In the simple-spin model, the photons dress the orbital state but not the spin state.  
We may therefore simply extend the longitudinal dressed states to include the bare spin:  $\ket{c;s;N}_l\equiv\ket{c;N}_l\ket{s}$.
In this basis, the diagonal elements of the full Hamiltonian are given by
\begin{equation}
E_N^{c,s}=N\hbar\omega -c E_{01}-s E_Z .  \label{eq:dressedspinE}
\end{equation}
Note that we have dropped the constant energy shift $g^2/\hbar\omega$ from Eq.~(\ref{eq:dressedE}) since it is present for all the basis states and plays no role in a following analysis.
In general, $\delta \hat{\mathbf B}$ is neither parallel nor perpendicular to ${\mathbf B}_\text{avg}$.
Hence, our spin-orbit coupling terms are of the form $\{\tilde{\sigma}_x,\tilde{\sigma}_z\}\otimes\{\tilde{\tau}_x,\tilde{\tau}_z\}$.
The off-diagonal perturbation in the fully quantum Hamiltonian is now given by
\begin{equation}
V=g\cos\theta\,\tilde\sigma_x(a+a^\dagger) 
+\sum_{j,k}^{x,z} f_{j,k}\,\tilde{\sigma}_j\tilde{\tau}_k ,
\label{eq:V}
\end{equation}
where the first term describes the orbital-photon coupling, and the second term describes the spin-orbit coupling.
The spin-orbit coupling constants $f_{j,k}$ depend, in part, on geometrical factors such as the relative alignment of the magnetic field gradient with respect to the dipole moment of the orbital states.
The term $f_{x,x}\,\tilde{\sigma}_x\tilde{\tau}_x$ is the conventional EDSR matrix element~\cite{Tokura2006}, describing the flip-flop transition between the orbital and spin states.
The constant $f_{z,x}$ is of particular interest for the resonance (5).
It describes a hybridization of spin states ($\tilde{\tau}_x$), which depends on the orbital occupation, but does not hybridize the orbital states ($\tilde{\sigma}_z$).
These coupling constants arise through the valley-orbit coupling mechanism, which depends on the local disorder realization at the quantum well interface.
Since we do not know the disorder potential for our device, we do not attempt to calculate $f_{j,k}$ here.
However, we provide estimates for several of these parameters below, based on our experimental measurements. 
The other contributions to the spin-orbit matrix elements can be evaluated using Eqs.~(\ref{eq:ME1})-(\ref{eq:ME3}), and the additional results
\begin{gather}
\hspace{-1.72in} _l\bra{c;N}\tilde\sigma_x\ket{c;N+m}_l = \nonumber \\ \hspace{0.2in}
_l\bra{-c;N}\tilde\sigma_z\ket{c;N+m}_l = 0 , \label{eq:ME4}  \\
\hspace{-1.27in} _l\bra{c;N}\tilde\sigma_z\ket{c;N+m}_l = 2c\,\delta_{m,0} , \label{eq:ME5} \\ 
\hspace{-1.6in} _l\bra{-c;N}\tilde\sigma_x\ket{c;N+m}_l \simeq \nonumber \\ \hspace{0.5in}
\left(2\alpha\sqrt{\langle N\rangle}\right)^{2m} \!\!\!
J_m\!\left(-4\alpha\sqrt{\langle N\rangle}\right)  , \label{eq:ME6}  
\end{gather}
which are also obtained using standard techniques~\cite{TannoudjiBook}.
Here, $J_m(x)$ is an integer Bessel function.

In the simple-spin model, no photons are absorbed via spin-orbit coupling, so only the $m=0$ terms will be considered in Eqs.~(\ref{eq:ME4})-(\ref{eq:ME6}).
There are therefore three types of direct (i.e, first-order) transitions allowed by Eq.~(\ref{eq:V}):
(i) PAT transitions, with no spin flip, like those described in Eq.~(\ref{eq:ME3}),
(ii) spontaneous flip-flops between the spin and orbital states ($f_{x,x}$ processes), 
or (iii) phase exchange ($f_{z,z}$) processes, without photon absorption or emission.
Note that direct transitions of the $f_{x,z}$ or $f_{z,x}$ type are also allowed; however, they do not conserve energy, and are therefore highly suppressed in resonant transitions.
The only relevant first-order matrix elements are therefore 
\begin{gather}
_l\bra{-c;s;N}V \ket{c;s;N+m}_l =  h\nu_{N,m}/2 \quad (m\geq 1) \label{eq:ME30} , \\
\hspace{-0.5in} _l\bra{c;s;N}V\ket{c;s;N}_l = 2cf_{z,z} , \label{eq:ME50} \\ 
\hspace{-0.in} _l\bra{-c;-s;N}V\ket{c;s;N}_l \simeq 
f_{x,x}\,J_0\!\left(4\alpha\sqrt{\langle N\rangle}\right)  . \label{eq:ME60}  
\end{gather}
Note that Eq.~(\ref{eq:ME30}) corresponds to PAT, while Eqs.~(\ref{eq:ME50}) and (\ref{eq:ME60}) describe pure spin-orbit coupling, which hybridizes the spin and orbital states, even in the absence of a driving term.

We now consider second-order processes in the perturbation, $V$.
All photon-driven processes with spin flips must be second-order, involving two distinct steps.
In the first, an orbital state is excited (deexcited) while absorbing (emitting) one or more photons; this is a nonresonant LZ process.
In the second, a spin flip occurs, accompanied by a second orbital flip (an $f_{x,x}$ process), or a phase change of the orbital state (an $f_{z,x}$ process).
Processes of the $f_{x,z}$ and $f_{z,z}$ type are also allowed; however they do not cause spin flips, and will not be considered in our EDSR analysis.
We first consider a pure-spin resonance generated by the sequence 
$\ket{c;s;N+m}\rightarrow\ket{-c;s;N}\rightarrow\ket{c;-s;N}$.
Here, the difference between the initial and final states involves a spin excitation and $m$ absorbed photons.
The intermediate state $\ket{-c;s;N}$ is off-resonant, and may be eliminated to obtain an effective, second-order matrix element, given by
\begin{multline}
_l\bra{c;-s;N}V\ket{c;s;N+m}_l \simeq \\
\frac{h\nu_{N,m}f_{x,x}}{4(sE_Z-cE_{01})}J_0\!\left(4\alpha\sqrt{\langle N\rangle}\right)  . \label{eq:ME7}
\end{multline}
Here, we have evaluated the energy denominator at the resonant condition $m\hbar\omega=2sE_Z$.
Similarly, we can consider resonances (5), (6), and (8) type process generated by the sequence
$\ket{c;s;N+m}\rightarrow\ket{-c;s;N}\rightarrow\ket{-c;-s;N}$.
In this case, the difference between the initial and final states involves a spin excitation, a charge excitation, and $m$ absorbed photons.
The effective, second-order matrix element for this process is given by
\begin{equation}
_l\bra{-c;-s;N}V\ket{c;s;N+m}_l \simeq 
-\frac{h\nu_{N,m}f_{z,x}c}{2sE_Z}, \label{eq:ME8}
\end{equation} 
corresponding to the resonance condition $m\hbar\omega=2(sE_Z+cE_{01})$.
In the limit $\alpha\sqrt{\langle N\rangle}\lesssim 1$, which seems appropriate for our experiments (see below), we therefore obtain
\begin{gather}
\hspace{0in}
_l\bra{c;-s;N}V\ket{c;s;N+m}_l \simeq
\frac{h\nu_{N,m}f_{x,x}}{4(sE_Z-cE_{01})} \equiv\frac{hf_m}{2} ,
\label{eq:Rabif} \\ \hspace{0in}
_l\bra{-c;-s;N}V\ket{c;s;N+m}_l \simeq 
-\frac{h\nu_{N,m}f_{z,x}c}{2sE_Z} \equiv\frac{hf_m'}{2} . \label{eq:Rabifp}
\end{gather}
Here, $f_m$ represents the Rabi frequency for a pure-spin resonance, and $f_m'$ represents the Rabi frequency for a combined spin-orbit resonance [like resonance (5)].

Keeping all other parameters fixed, we can estimate the power scaling of the Rabi frequencies (\ref{eq:Rabif}) and (\ref{eq:Rabifp}) by using Eqs.~(\ref{eq:gcouple}), (\ref{eq:alpha}), and (\ref{eq:nulargeN}).
Assuming the limit $\alpha\sqrt{\langle N\rangle}\rightarrow 0$, we find that $f_1,f_1'\propto \epsilon_1$, $f_2,f_2'\propto\epsilon_1^2$, etc.; these theoretical results are consistent with experimental results reported in Ref.~\cite{Scarlino2015}.
Based on Eq.~(\ref{eq:Rabif}), we also predict a simple form for the ratio 
\begin{equation}
\frac{f_2}{f_1}\simeq\frac{\nu_{N,2}}{\nu_{N,1}}\simeq -2\alpha\sqrt{\langle N\rangle} .
\end{equation}
Using the results for $f_1$ and $f_2$ in Ref.~\cite{Scarlino2015}, we estimate this ratio to be in the range $2\alpha\sqrt{\langle N\rangle}\simeq 1/3$-1/2 for typical experiments, which confirms our previous assumptions. For the experiments reported here, we use lower microwave power than in  Ref.~\cite{Scarlino2015}, which yields a smaller estimate for this parameter, but makes it impossible to measure $f_1$ and $f_2$.

We can estimate the ratio of the Rabi frequencies between different orbital states (i.e., different values of $c$), but the same EDSR mode (the same value of $m$).
Taking $c=\pm 1/2$ in Eq.~(\ref{eq:Rabif}), we obtain the ratio
\begin{equation}
\frac{f_m^{(1)}}{f_m^{(2)}}\simeq\frac{E_Z+E_{01}}{E_Z-E_{01}} . \label{eq:Rabiratio}
\end{equation}
Here, we have adopted the notation of Ref.~\cite{Scarlino2015} where $f_1^{(1)}$ ($f_1^{(2)}$) corresponds to $c=1/2$ ($c=-1/2$) for the case $m=1$.
In Ref.~\cite{Scarlino2015}, it was found that $f_1^{(1)}/f_1^{(2)}=1.7$ when $B_\text{ext}=561$~mT.
Using these values, Eq.~(\ref{eq:Rabiratio}) predicts that $E_{01}=17$~$\mu$eV or 4~GHz.
We note that, although Ref.~\cite{Scarlino2015} used the same device as we do here, the gates were tuned differently, which could potentially yield to different values for the valley splitting. The spectroscopy shown in Fig.~1 of the main text gives a more direct estimate of  $E_{01}=7$~GHz.
The agreement is reasonable; however, Eq.~(\ref{eq:Rabiratio}) also predicts that $f_m^{(1)}/f_m^{(2)}$ should not depend on $m$, while Ref.~\cite{Scarlino2015} estimates a ratio of $f_2^{(1)}/f_2^{(2)}=0.9$ for the first subharmonic resonance ($f_1^{(3)}/f_1^{(4)}$ in their notation).
This experimental result is clearly inconsistent with Eq.~(\ref{eq:Rabiratio}), which should be greater than 1.
We point out, however, that the error bars reported in Ref.~\cite{Scarlino2015} were large enough to permit $f_2^{(1)}/f_2^{(2)}>1$. Moreover, we note again that strong driving was employed in Ref.~\cite{Scarlino2015}, so that the arguments leading to  Eq. (S37) could begin to break down.

The scaling behavior $f_1\sim f_{x,x}\epsilon_1/(E_Z-E_{01})$ suggested by Eq.~(\ref{eq:Rabif}) is consistent with Tokura et al.~\cite{Tokura2006}.
We can now obtain a rough estimate for the EDSR coupling constant $f_{x,x}$ by combining the current experimental results with those reported in Ref.~\cite{Scarlino2015}.
In Ref.~\cite{Scarlino2015}, a typical value of $f_1^{(1)}=2.5$~MHz was obtained (note that $f_1^{(1)}$ should be smaller in the current experiment, although we do not know its value).  In the current experiments, we find that $E_Z-E_{01}\simeq 7$~GHz in the range of interest (see Fig.~1), and $\nu_{N,1}\simeq 0.2$~GHz (see Sec.~S.II.C, below).
From Eq.~(\ref{eq:Rabif}), we then obtain the estimate $f_{x,x}\simeq 90$~MHz.
We can also estimate the parameter $f_{z,x}$.
As discussed above, this coupling constant describes the orbital (i.e., valley)-induced hybridization of the spin states, and determines the strength of the resonances (5) and (6). Specifically, it arises from contributions to the $g$-factor tensor that are transverse to the applied field, and which differ slightly between the two different valley states.
It is not possible to compute these valley-induced perturbations without knowledge of the disorder potential.
In general however, the disorder will not be aligned with the magnetic field, so the transverse component of the perturbation in the $g$-factor tensor should differ from the total perturbation by a simple geometrical factor of order $O[1]$.
In Ref.~\cite{Kawakami2014}, the total difference in the $g$-factors for the two valley states was found to be $\Delta g/g\simeq 0.015$\%.
Hence, we can estimate the valley-dependent hybridization of the spin states to be $2f_{z,x}=\Delta g\mu_BB$. At the magnetic field corresponding to the kink in hybridized resonance (5),(6) in Fig.~1 of the main text, we have $B_\text{tot}\simeq 480$~mT, yielding $f_{z,x}\simeq 1$~MHz, or $f_{z,x}/f_{x,x}\simeq 1$\%.
Finally, we can estimate a typical Rabi frequency for resonance (5) from Eq.~(\ref{eq:Rabifp}), giving $f_1'\simeq 10$~kHz.
This small value explains why it is not possible to observe coherent oscillations of the resonance (5) in our experiments.

To close this section, we discuss the basic features of the simulation results shown in Fig.~\ref{fig:theorylabels}, in terms of the first and second-order matrix elements obtained in Eqs.~(\ref{eq:ME30})-(\ref{eq:ME60}) and (\ref{eq:Rabif})-(\ref{eq:Rabifp}).
First, it is helpful to recall how the spin enters this discussion, since the orbital Hamiltonian used in the simulations, Eq.~(\ref{eq:H}), does not explicitly include spin.
It is important to note that Fig.~\ref{fig:theorylabels} does not describe spin resonance phenomena; it is simply the Fourier transform of the response of the orbital dipole moment to an ac drive.  
In the simple-spin model, this response is converted into an ac magnetic signal, with a strength determined by the spin-orbit coupling parameters $f_{j,k}$ in Eqs.~(\ref{eq:ME50}) and (\ref{eq:ME60}), yielding an EDSR response in the spin sector.
Since there is no spin-orbit coupling in the simulations, we can learn nothing about the spin Rabi frequencies.
However, the response spectrum in Fig.~\ref{fig:theorylabels} should still provide an accurate mapping of the EDSR spectrum.

\begin{figure*}[ht!]
\includegraphics[width=5.5in]{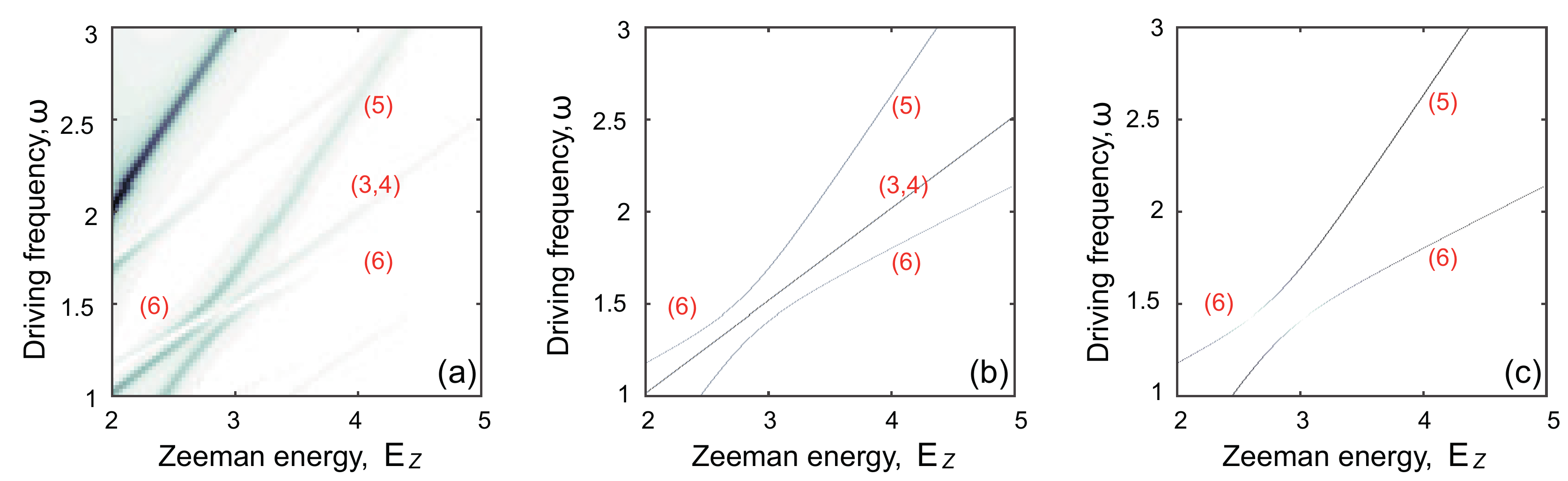}
\caption{
Two types of theory estimates for the EDSR resonance spectra.
(a) Simulation results, reproduced from Fig.~\ref{fig:theorylabels}, with the resonances of interest labeled.
To make contact with the dressed-state theory, we have equated the response frequency $\Omega$ to the Zeeman energy $E_Z$.
(b) Spin-state resonances obtained from the 4D dressed-state Hamiltonian, Eq.~(\ref{eq:H4Dsimple}).  
Here, we use $E_{01}=\sqrt{2}$ and $h\nu_1=0.22$. 
(c) Orbital-state resonances obtained for the same parameters.
}
\label{fig:quad} 
\end{figure*}

To take some examples, the features label (1),(2) and (3),(4) in Fig.~\ref{fig:theorylabels} correspond to conventional EDSR, with spin excitations occurring at the resonance conditions $E_{N+m}^{c,s}=E_N^{c,-s}$ (or $m\hbar\omega=2sE_Z$) when $m=1$ and 2, respectively.
These features correspond to the experimenal resonances labeled the same way in Fig.~1(a) of the main text.
The corresponding Rabi frequencies are given by $f_m$ in Eq.~(\ref{eq:Rabif}), and were measured to be in the range of a few MHz in Ref.~\cite{Scarlino2015}.
The feature labeled (9),(10) in Fig.~\ref{fig:theorylabels} corresponds to a 3-photon spin resonance, with $m=3$; this feature is not observed in the experiments.
The features labeled (5), (6), and (8) occur at the resonance conditions $E_{N+m}^{c,s}=E_N^{-c,-s}$ [or $m\hbar\omega=2(sE_Z+cE_{01})$], and involve simultaneous spin and orbital excitations.
In previous work, such processes were referred to as ``spin-flip PAT"~\cite{Schreiber2011,Braakman2014}.
The corresponding Rabi frequencies are given by $f_m'$ in Eq.~(\ref{eq:Rabifp}), and have not been measured yet experimentally.
In principle, the PAT transitions, which occur at the resonance condition $E_{N+m}^{c,s}=E_N^{-c,s}$ (or $m\hbar\omega=2cE_{01}$) in Eq.~(\ref{eq:ME30}), could also be observed in the simulations.
However, these transitions correspond to discrete points along the line labeled (7) in Fig.~\ref{fig:theorylabels}, which makes them difficult to detect.  
Moreover, the response frequency for PAT ($E_{01}/h$) is smaller than Fermi level broadening, so it cannot be detected in our readout scheme.
It is therefore interesting that PAT plays a key role in our understanding of the hybridization between dressed states, as discussed in Sec.~S.II.C, and that the repulsion between the dynamical modes allows us to estimate the magnitude of the PAT Rabi frequency, $\nu_{N,m}$.

One of the most interesting features in Fig.~\ref{fig:theorylabels} is labeled (7), and corresponds to a pure-LZ transition, with an orbital state excitation but no photon absorption.
Since $m=0$, according to Eq.~(\ref{eq:ME3}), this cannot be described by a direct, first-order process.  
Instead, it involves a combination of photon-mediated processes, with a \textit{net} photon absorption of $m=0$. 
A minimal process of this type must be third-order, since second-order processes would involve an orbital excitation, followed by a deexcitation, leaving the orbital in its initial state.
Since the electrons in experiments have spins (in contrast with our simulations), there is also another mechanism that could cause feature (7) in the experiments: a spin-orbit flip-flop occurring at the resonance condition $E_N^{c,s}=E_N^{-c,-s}$ (or $sE_Z=-cE_{01}$). This is the direct, first-order process, described by Eq.~(\ref{eq:ME60}), in which the orbital state spontaneously deexcites while causing a spin excitation.
Feature (7) should therefore appear in both the simulations and the experiments, but with different physical origins and different Rabi frequencies.
This is a moot point for experiments however, since low-energy processes cannot be resolved by our measurement scheme.

\subsection{C. Hybridization of the Dressed States} \label{sec:hybrid}

Dressed states allow us to analyze one of the most intriguing features of the data: the kink in the hybridized resonance (5),(6), where the resonance line appears to switch between two well-defined slopes, as shown in Fig.~1(b) of the main text.
The two asymptotic resonance lines correspond to the conditions $E_{N+3}^{+1/2,+1/2}\simeq E_N^{-1/2,-1/2}$ (or $3\hbar\omega\simeq E_Z+E_{01}$) far to the left of the kink, and $E_{N'+1}^{-1/2,+1/2}\simeq E_{N'}^{+1/2,-1/2}$ (or $\hbar\omega\simeq E_Z-E_{01}$) far to the right.
The intersection of these two resonance lines occurs when $E_Z=2\hbar\omega=2E_{01}$, or $N'=N+1$, and defines a quadruple resonance. At this special point, the arrows in the processes labeled (3)-(7) in Fig.~1(d) all have the same length.
The Hamiltonian for this 4D manifold is given by
\begin{widetext}
\begin{equation}
H_\text{4D}=\begin{pmatrix}E_N^{-1/2,-1/2} & \frac{h\nu_{N,1}}{2} & \frac{hf_2}{2} & \frac{hf_3'}{2} \\
\frac{h\nu_{N,1}}{2} & E_{N+1}^{+1/2,-1/2} & \frac{hf_1'}{2} & \frac{hf_2}{2} \\
\frac{hf_2}{2} & \frac{hf_1'}{2} & E_{N+2}^{-1/2,+1/2} & \frac{h\nu_{N+2,1}}{2} \\
\frac{hf_3'}{2} & \frac{hf_2}{2} & \frac{h\nu_{N+2,1}}{2} & E_{N+3}^{+1/2,+1/2} 
\end{pmatrix} . \label{eq:H4D}
\end{equation}

We can gain intuition about the experiments and simulations by considering a simplified version of $H_\text{4D}$.
First, we note that $\nu_{N,1}\simeq\nu_{N+2,1}\equiv\nu_1$ when $\langle N\rangle\gg 1$.
To investigate the dominant behavior, since $\nu_1 \gg f_m , f_m'$, we consider the limit $f_m,f_m'=0$.
The resulting Hamiltonian
\begin{equation}
H_\text{4D}\simeq\begin{pmatrix}E_{01}+E_Z & \frac{h\nu_{1}}{2} & 0 & 0 \\
\frac{h\nu_{1}}{2} & \hbar\omega+E_Z & 0 & 0 \\
0 & 0 & 2\hbar\omega+E_{01} & \frac{h\nu_{1}}{2} \\
0 & 0 & \frac{h\nu_{1}}{2} & 3\hbar\omega 
\end{pmatrix} -\frac{E_{01}+E_Z}{2} \label{eq:H4Dsimple}
\end{equation}
\end{widetext}
does not permit any dynamical hybridization of spin states, and can therefore be used to directly identify the spin resonance conditions.

We find that Eq.~(\ref{eq:H4Dsimple}) reproduces all the main features of the experimental and simulation results near the kink in hybridized resonance (5),(6), as indicated in Fig.~\ref{fig:quad}.
The features observed in Figs.~\ref{fig:quad}(b) and (c) reflect sudden changes in the compositions of the eigenstates.
To obtain these plots, we first diagonalize $H_\text{4D}$ to obtain its four eigenstates as a function of $E_Z$.
We then compute probability $P^{(j)}(s)$ for the $j$th state to have spin $s$.
Here, the states are labeled according to their energy ordering, as they would be for eigenstates of the full 4D Hamiltonian in Eq.~(\ref{eq:H4D}).
Near a spin resonance, the spin-up or spin-down character of the ordered eigenstates changes abruptly.
To see these sudden changes in the simulations, we sum the numerical derivatives, $\sum_j|\partial P^{(j)}(s)/\partial E_Z|$.
The resulting spectrum displays sharp peaks at the resonance locations, as illustrated in Fig.~\ref{fig:quad}(b).
Comparison with the simulation results, reproduced in Fig.~\ref{fig:quad}(a), shows that only one parameter ($h\nu_1/E_{01}\simeq 0.16$) is needed to capture the main features of the hybridized resonance (5),(6).
Including the other off-diagonal terms in Eq.~(\ref{eq:H4D}) leads to small feature changes, such as broadening of the resonance peaks, but does not change the underlying spectrum.

The dynamically induced shifts of the dressed state energies (i.e., a.c. Stark shifts) are dominated by the orbital coupling terms $\nu_{N,m}$, because $\nu_{N,m}\gg f_m$.
Hence, they are easily computed from Eq.~(\ref{eq:H4Dsimple}) by diagonalizing the two 2D subsystems.
Diagonalizing the Hamiltonian in this way yields the four energies
\begin{gather}
\frac{E_Z+\hbar\omega}{2}\pm\sqrt{\left(\frac{E_{01}-\hbar\omega}{2}\right)^2+\left(\frac{h\nu_1} {2}\right)^2}
\label{eq:acStark1},\\
\frac{5\hbar\omega-E_Z}{2}\pm\sqrt{\left(\frac{E_{01}-\hbar\omega}{2}\right)^2+\left(\frac{h\nu_1} {2}\right)^2}.
\label{eq:acStark2}
\end{gather}
In particular, the energy splitting between hybridized resonance (5),(6) and its large-$E_Z$ asymptote, $\hbar\omega=E_Z-E_{01}$, is given by
\begin{equation}
\Delta E_\text{asymp} \simeq \frac{(h\nu_1)^2}{2(E_Z-2E_{01})} ,
\end{equation}
and the energy splitting at the anticrossing between resonances (3),(4) and (5),(6) is given by $\Delta E_X=h\nu_1/\sqrt{3}$.
Using the asymptotic formula for the Rabi frequencies, Eq.~(\ref{eq:nulimit}), and in particular, $\nu_1\simeq\epsilon_1\cos\theta/2h$, we find that $\Delta E_X\simeq \epsilon_1\cos\theta/2\sqrt{3}$ in the regime of experimental interest.
We can also obtain analytical expressions for the resonance conditions by equating the energy expressions in Eqs.~(\ref{eq:acStark1}) and (\ref{eq:acStark2}).
For the resonant feature labeled (3),(4) in Fig.~\ref{fig:quad}, we obtain the condition $\hbar\omega=E_Z/2$. 
For (5),(6), we obtain the conditions
\begin{equation}
\hbar\omega =  \tfrac{1}{3}(2E_Z-E_{01})\pm \tfrac{1}{3}\sqrt{(E_Z-2E_{01})^2+3(h\nu_1)^2} .
\end{equation}
Note that this equation describes the hybridized resonances on either side of (3),(4).

It is interesting to note that, while the parameters $f_m$ control the spin dynamics and the EDSR Rabi frequencies, the dressed state energies $E_N^{c,s}$ determine the gross features of the resonance spectrum, and the PAT Rabi frequency $\nu_1$ determines the fine features of hybridized resonance (5),(6).
From Eqs.~(\ref{eq:H4D}) and (\ref{eq:H4Dsimple}), we see that resonances (5) and (6) correspond to dressed states with the same spin but different orbital states.  
The kink in the hybridized resonance is therefore caused by orbital physics. The correspondence between Figs.~\ref{fig:quad}(a) and \ref{fig:quad}(b) indicates that all of the resonances in this manifold involve a spin flip.
We have also computed the analogous orbital resonances for the same 4D manifold, corresponding to sudden changes in the charge state, as obtained from the expression $\sum_j|\partial P^{(j)}(c)/\partial E_Z|$.
The results are shown in Fig.~\ref{fig:quad}(c).
In this case, the central resonance (3),(4) disappears, indicating that it is a pure-spin resonance. 

Although coherent EDSR oscillations were observed in Ref.~\cite{Scarlino2015}, with Rabi frequencies in the range $f_1,f_2\simeq 0.6$-4~MHz, it was not possible to observe coherent PAT oscillations because of their relatively low energy scale.
However, we can now estimate the PAT Rabi frequency by fitting the analytical theory of Eq.~(\ref{eq:H4Dsimple}) to the experimental resonance spectrum in Fig.~1(b) of the main text, obtaining the result $\nu_1\simeq 0.2$~GHz.
This confirms our previous assumption that $f_1\ll\nu_1$.
We can also estimate the remaining parameters in the theoretical model of Eqs.~(\ref{eq:H}) and (\ref{eq:epst}).
From Eqs.~(\ref{eq:gcouple}) and (\ref{eq:nulargeN}), we obtain $\epsilon_1\simeq 0.5$~GHz~$\simeq 2$~$\mu$eV.
Using the experimental result $E_{01}\simeq 7$~GHz~$\simeq 29$~$\mu$eV and the rough estimate $\epsilon_0\simeq\Delta$ (deduced from our simulations), we also obtain $\epsilon_0\simeq\Delta\simeq E_{01}/\sqrt{2}\simeq 5$~GHz~$\simeq 20$~$\mu$eV.
The parameters $\epsilon_0$ and $\epsilon_1$ that describe the detuning of the orbital basis states due to the driving field are difficult to determine from first principles in silicon devices because they are determined by the degree of valley-orbit mixing, which depends sensitively on the details of the interfacial disorder~\cite{Gamble2013,Kharche2007}.  In contrast, the energy splitting between the orbital states $E_{01}$ and the PAT frequency $\nu_1$ can be extracted directly from the experiments.

\section{S.III. ADDITIONAL MEASUREMENTS AND ANALYSIS} \label{sec:additional measurements}

\begin{figure} 
\includegraphics[width=4.5cm]{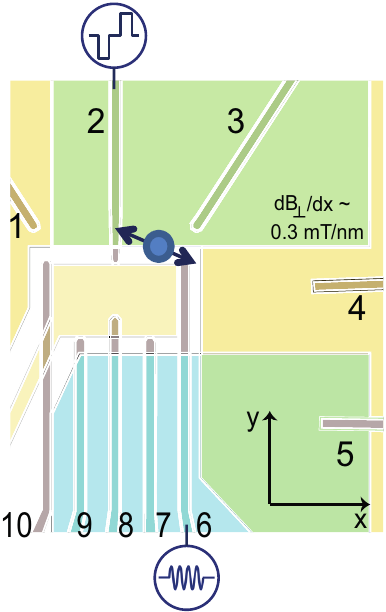}
\caption{\label{fig:figS1} Schematic of the device, showing a pattern of split gates, labeled 1-10. For this experiment we create a single quantum dot (estimated location indicated by a blue circle) and a sensing dot (gates 4 and 5). The current through the sensing dot is recorded in real time for a fixed voltage bias of 500 $\mu$eV. The voltage pulses and microwave excitation are applied to gate 2 and 6 respectively. Green semitransparent rectangles show the position of two 200-nm-thick Co micromagnets. The yellow-shaded areas show the location of two accumulation gates, one for the reservoirs and the other for the double quantum dot region.} 
\end{figure}

\begin{figure*}[ht!]
\centering
\includegraphics[width=0.7\linewidth]{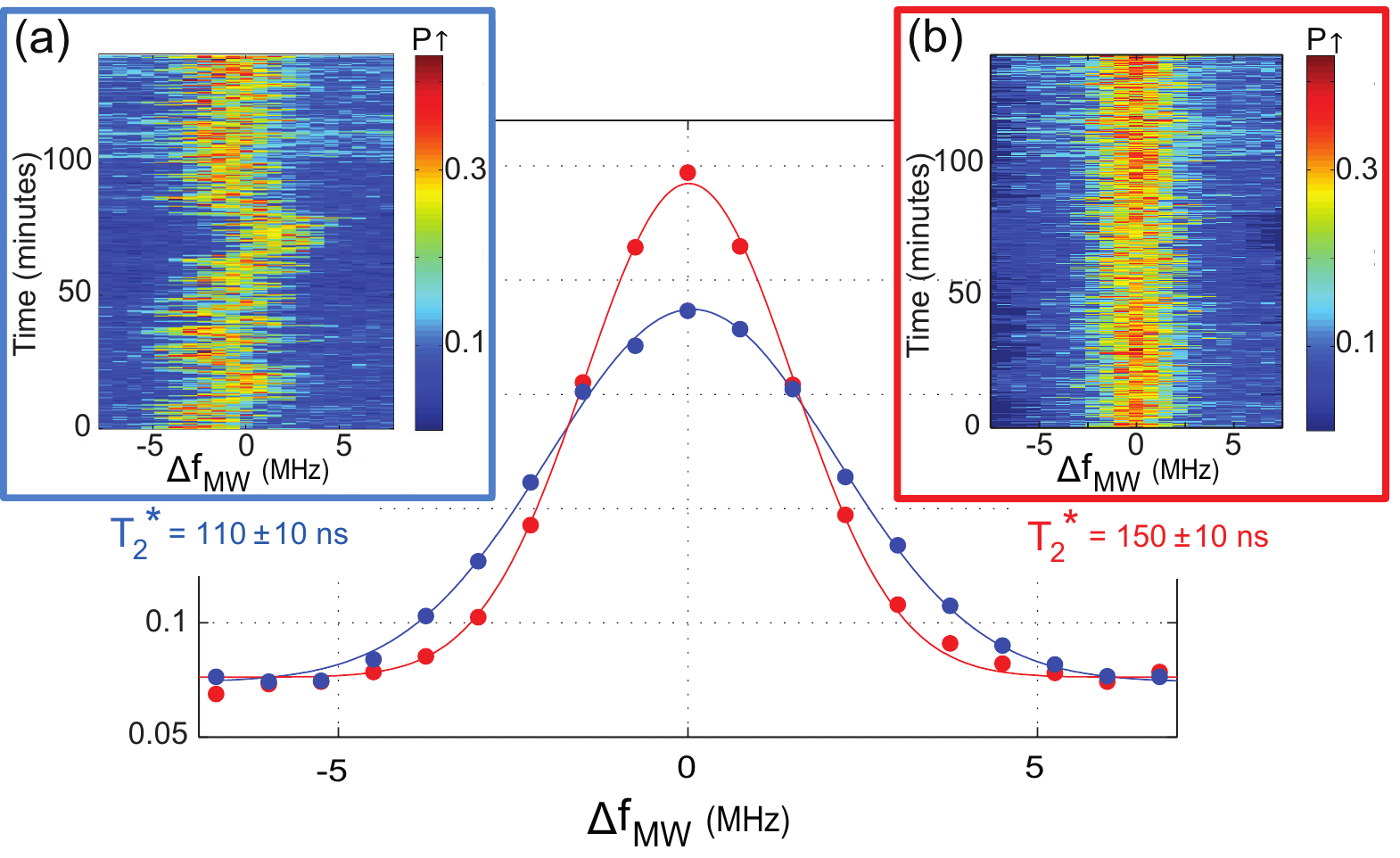}
\caption{The main panel represents a CW (low MW power) measurement of the inter-valley spin resonance, done for $B^y_{ext}=860$ mT. The blue and red traces are obtained averaging over time the data reported respectively in panels (a) and (b). Both panels report the CW resonance, away from 18.223 GHz, recorded over time ($y$ axis). Each point is averaged over 10 ms and there are 150 points per trace. We observe that over time the resonance shifts in frequency. Averaging the bare data in panel (a) we get the blue trace. Fitting with a Gaussian we can get a lower bound for the $T_2^\ast$ of $\sim$110 ns. Furthermore, we can operate a different kind of averaging procedure: we can fit each trace with a Gaussian, get the center of each fit, shift each set of data to center all of the resonance traces [see panel (b)] and average all of them. In this way we get the red trace in the main panel, that gives us a $T_2^\ast$ of $\sim$150 ns. In the first method, we average on a timescale of a few hours, in the second on a timescale of tens of seconds only. The difference between the two average procedures points at the importance of very-low-frequency noise.}
\label{fig:Fig3}
\end{figure*}

\begin{figure*}[ht!]
\centering
\includegraphics[width=0.7\linewidth]{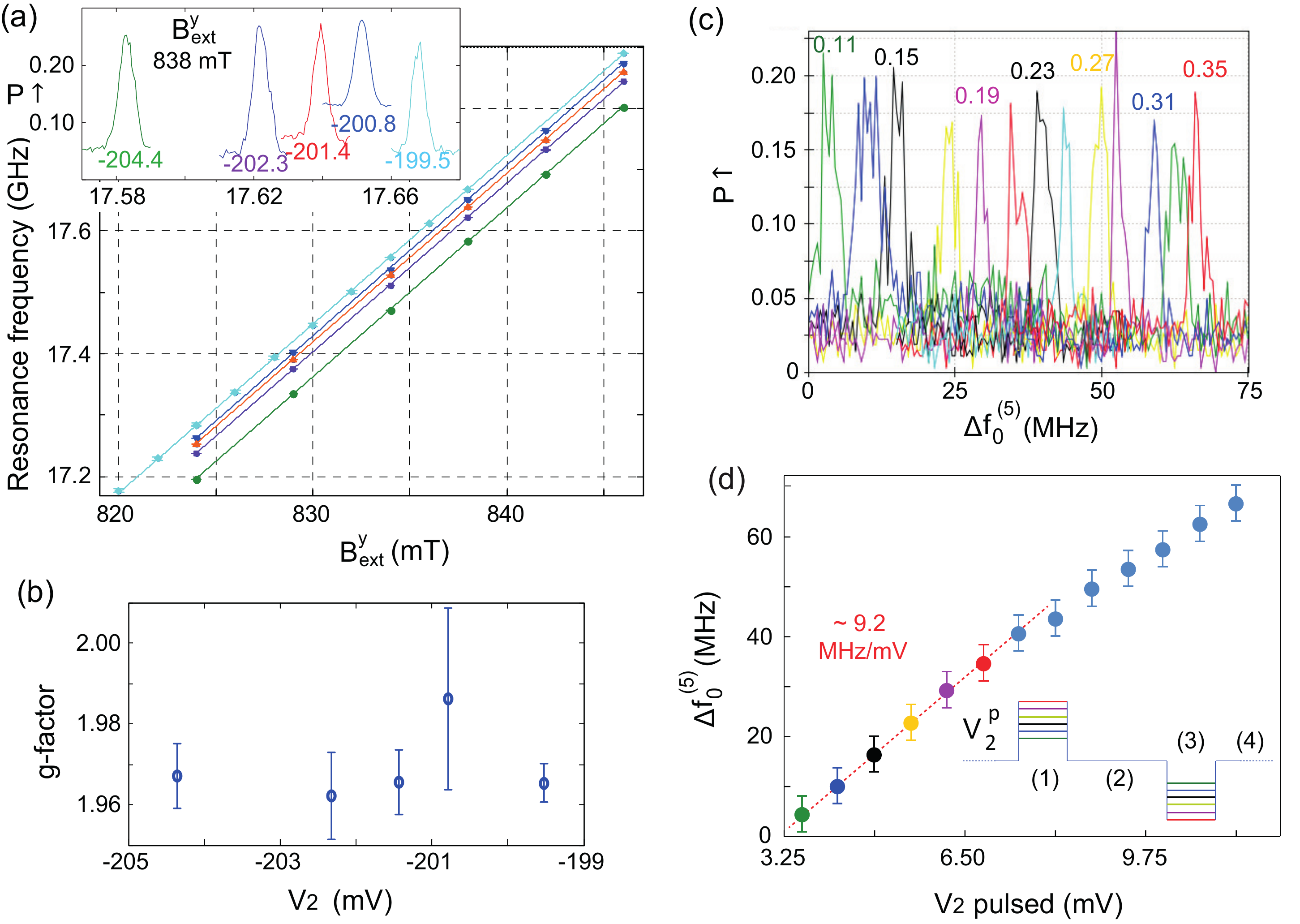}
\caption{(a) $g$-factor measurements for different voltage applied to gate 2 ($V_2$), reported in mV with different colors in the inset; the extracted $g$-factors for each value of $V_2$ are reported in panel (b). The fact that the g-factor stays constant suggests that the mechanism responsible for the gate dependence of the Larmor frequency $f_0^{(5)}$ is not related to the shift of the electron wave function in the micromagnet stray field (see main text). In panel (c) we show that it is possible to shift the resonance frequency also by changing the voltage amplitude during the manipulation stage. We keep the pulse shape symmetric, compensating by using the same voltage pulse amplitude, but with opposite sign, during the initialization stage [see inset of panel (d)]. The number reported on each resonance represents the voltage pulse amplitude (in V) at the output of the AWG. (d) Shift in the resonance frequency $f_0^{(5)}$ as a function of a gate voltage pulse applied to gate $V_2$ [the color code of the 6 voltage pulse levels corresponds to the first 6 resonances reported in panel (c)]. Here, $\Delta f_0^{(5)}= f_0^{(5)}-21$ GHz and for the $x$ axis we used the conversion factor 32.5 mV/V extracted from calibration measurements.}
\label{fig:Fig4}
\end{figure*}

\begin{figure*}[ht!]
\centering
\includegraphics[width=0.7\linewidth]{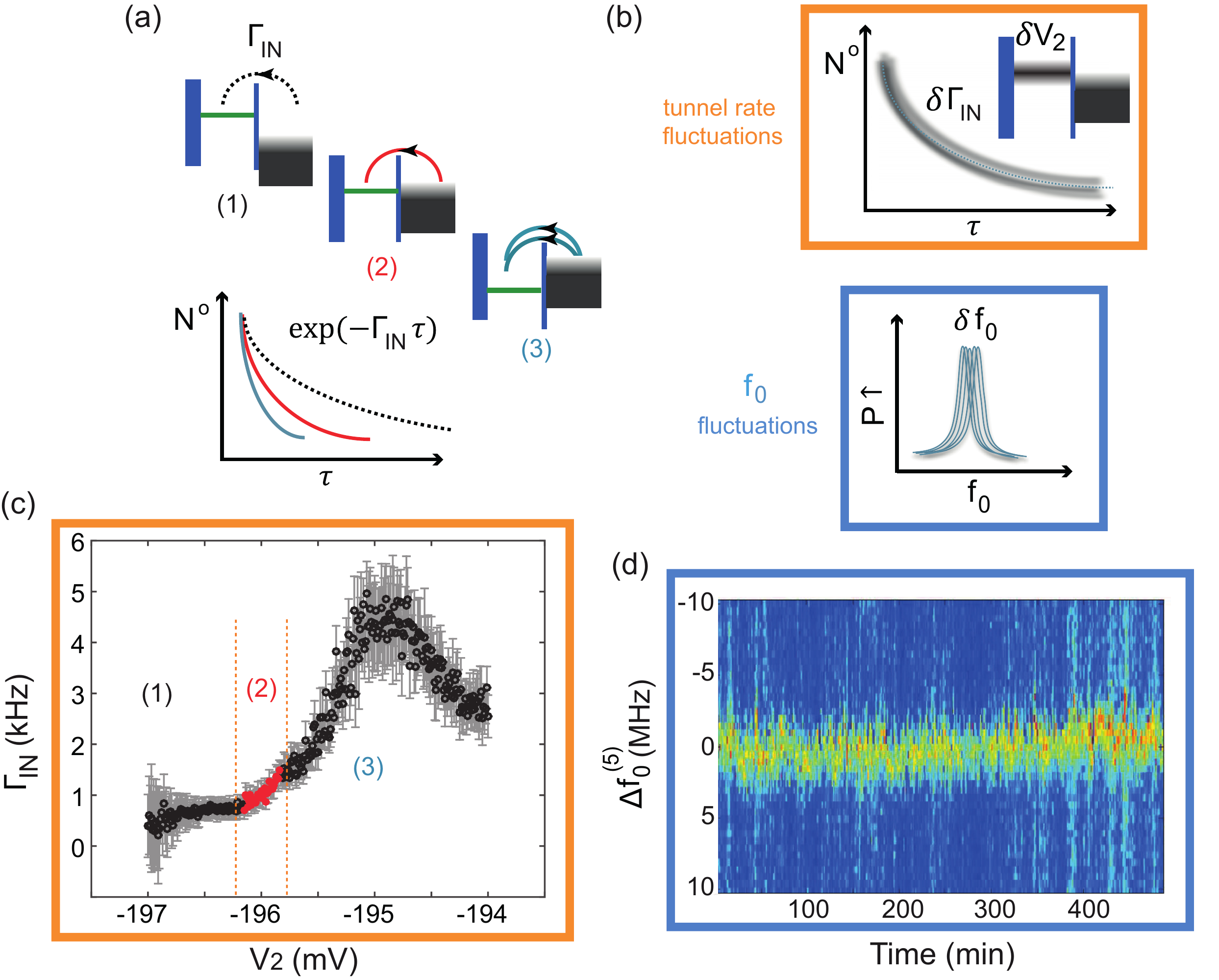}
\caption{\label{fig:figS4}(a) Schematic representing the variation of the tunnel-in rate $\Gamma_\text{IN}$, as a function of the relative detuning between the electrochemical potential in the dot and the Fermi level in the reservoir, controlled by $V_2$. (b) Schematic of the effect of the magnetic noise and electrical noise, causing respectively a Larmor frequency fluctuation (lower panel) and tunnel rate fluctuations (upper panel).
(c) Tunnel-IN rate, $\Gamma_\text{IN}$, recorded during the initialization stage as a function of $V_2$. We are operating around the readout position indicated by the red points in panel (c) (for which we can estimate a sensitivity of $\frac{d \Gamma_\text{IN}}{d V_2}\sim$1.7 kHz/mV). (d) CW measurement of the inter-valley resonance at low MW power (as in Fig.~\ref{fig:Fig3}) over time, in order to study the low frequency (tens of mHz) noise spectrum affecting the inter-valley spin resonance frequency. ($B^x_{ext}=590$ mT; $B^y_{ext}=598.2$ mT).}
\end{figure*}

\subsection{A. Experimental methods}\label{sec:Experimental details7}

All the measurements shown here make use of 4-stage voltage pulses [[see inset of Fig.~4(a)] applied to gate 2 \cite{Kawakami2014} (see Fig.~\ref{fig:figS1}): (1) initialization to spin-down (4 ms), (2) spin manipulation through microwave excitation of gate 8 (1 ms), (3) single-shot spin read-out (4 ms), and (4) a compensation/empty stage (1 ms). By repeating this cycle, typically 150-1000 times, we collect statistics of how often an electron leaves the dot during the detection stage, giving the spin-excited state probability, $P_\uparrow$, at the end of the manipulation stage.

Microwave excitation, generated by an Agilent E8267D Vector Source, is applied to gate 6 through the following attenuation chain: DC block/ULT-05 Keycom coax/20 dB attenuator (1.7 K)/ NbTi coax /10 dB attenuator (50 mK)/flexible coax/SMA connector on the printed circuit board (PCB). The high frequency signal is combined on the PCB with a DC voltage line by a homemade resistive bias tee ($R=10$ M$\Omega$, C= 47 nF).

\subsection{B. Variations in time of the spin-valley transition frequency}

In Fig.~\ref{fig:Fig3} we report repeated measurements of resonance (5) recorded over time, in order to investigate the low frequency noise fluctuations. In the main central panel we report, in blue, the spin resonance peak obtained by averaging directly, over time, the data of panel (a), inside the blue square frame. From its full width at half maximum (FWHM) we can get a lower bound for $T_2^\ast$, in this case of 100 ns. Alternatively we can also perform a Gaussian fit for each of the time traces in panel (a) and then shift the center of each dataset in order to align the centers of each resonance peak on top of each other, as reported in panel (b). If we now perform the averaging we get the red Gaussian, reported in the main central panel (a), from which we can extract a $T_2^\ast$ of 150 ns. This last averaging strategy is equivalent to filtering out the very low frequency component of the noise ($\sim$ mHz), clearly visible in panel (a)~\cite{NoteS4}.\\

In what follows we report three different measurements realized in order to investigate the sensitivity of resonance (5) to the electrostatic environment and to electrical noise.

\subsection{C. Influence of the static and pulsed gate voltages}\label{sec:Influence of the static electric field7}

Controlling the electrostatic potential by the voltage applied to the gates defining the QD confinement potential, we can systematically shift in frequency resonance (5) (see Fig.~3). 
The inset of Fig.~\ref{fig:Fig4}(a) shows the peak of resonance (5) for five different voltage configurations of the gate 2 ($V_2$, represented by different colors).
The main panel reports a measurement of the g-factor in each of the gate voltage configurations reported in the lower inset of Fig.~\ref{fig:Fig4}(a). The extracted $g$-factors are compatible with each other, considering their associated errors, as reported in Fig.~\ref{fig:Fig4}(b). This implies that, by changing the gate voltage we can modify the Larmor frequency of the 2-level system involved in the resonance process, without appreciably modifying the Zeeman energy~\cite{NoteS5}. This suggests that, in this case, we are directly modifying the valley splitting [modelled by moving the QD towards a step defect in the QW, as schematically represented in Fig.~3(b)]. 

We can estimate the effective wave function displacement in space (mainly in the QW plane) generated by modifying the voltage applied to the gate 2 by 1 mV, by using the shift of the Larmor frequency of the intra-valley spin resonance [red trace in Fig.~3(a)] due to the magnetic field gradient. 
From the simulation of the magnetic field gradient we can estimate an upper bound for the magnetic field gradient of $\sim$0.2 mT/nm (in the $y$ direction $\sim$gate 2), equivalent to $\sim$56 MHz/nm in silicon. This energy gradient gives a position lever-arm of $\sim$0.01 nm/mV, considering the slope of $\sim$0.55 MHz/mV of the red curve in Fig.~3(a). 
Furthermore, by making use of the measured value 18.5 MHz/mV for $f_0^{(5)}$ [see blue trace in Fig.~3(a)] and of the estimation above $\sim0.01$ nm/mV, we can conclude that $\Delta E_{01}/\Delta|\textbf{r}|\sim 1.8$ GHz/nm (where $\textbf{r}$ is the vector representing the motion of the electron generated in the Si QW by the changing of the voltage on gate 2)~\cite{NoteS6}.

An equivalent way to explore this effect is reported in Figs.~\ref{fig:Fig4}(c) and \ref{fig:Fig4}(d)  where keeping a constant static d.c.~voltage configuration, we change the amplitude of the voltage pulse applied on gate 2 to bring the system in Coulomb Blockade during the manipulation stage. As summarized in Fig.~\ref{fig:Fig4}(d), by increasing the pulse amplitude $V^p_2$, the resonance peak systematically moves toward lower frequencies. The lever-arm of this process ($\Delta f /\Delta V^p_2$) is around 9 MHz/mV.
This measurement demonstrates the dynamic Stark shifting of resonance (5) and creates opportunities for site-selective addressing (voltage pulse induced addressability). 
The difference in $\Delta f/\Delta V_2$ between the d.c.~and pulsed case can be ascribed to the fact that for the measurement reported in Fig.~3(a) we have to compensate the change in $V_2$ by changing the voltage on another gate, in order to keep a good initialization-readout fidelity. Instead, for the pulsed case we just modify the voltage pulse amplitude on the gate 2 in the manipulation stage (and the empty stage, to keep the symmetry of the pulse), without any further compensation.

\subsection{D. Influence of electric field noise} \label{sec:correlation}
The measurements reported above reveal that the electrostatic configuration considerably affects the inter-valley resonance frequency. Next we consider the effect of the electrical noise on the coherence of the inter-valley spin resonance.

We already have clear evidence of the importance of the electric noise from the upper bound of $T_2^\ast$ of resonance (5), estimated from the FWHM measurement reported in in Fig.~\ref{fig:Fig3}, which is 10 times shorter than what it should be if just hyperfine fluctuations dominated \cite{Kawakami2014}.
This results from the higher sensitivity of resonance (5) to the electrical environment, so also to the electrical noise. 
In the hypothesis that the hyperfine (hf) and the electric noise (el) are independent, we can write the total noise affecting this resonance as $\delta S_\text{tot}=\sqrt{(\delta S_\text{hf})^2+(\delta S_\text{el})^2}$, from which\\
$[T_2^\ast]_\text{el}\sim 1/\sqrt{([T_2^\ast]_\text{tot})^{-2}-([T_2^\ast]_\text{hf})^{-2}}\sim 100$ ns. Therefore, we can deduce that the coherence properties of the inter-valley spin resonance are almost completely dominated by the electric noise (here for $[T_2^\ast]_\text{hf}$ we use the value 950 ns extracted for the inter-valley spin resonance in \cite{Kawakami2014}).

Now, we present a measurement [reported in Fig.~3(d)] realized to study the correlation between the low frequency fluctuations of resonance (5) and low frequency electric noise (low with respect to the electron tunnel rate during the readout stage).
In this respect we made use of the following measurement technique (see Fig.~\ref{fig:figS4}):\\
(i) the tunnel rate for an electron jumping inside (or outside) a QD from (into) an electron reservoir is a sensitive function of the relative energy alignment between the Fermi level of the electron reservoir and the electrochemical potential of the single electron inside the QD \cite{Vandersypen2004}. Fig.~\ref{fig:figS4}(a) reports a schematic representation of $\Gamma_\text{IN}$ (the single electron tunnel rate IN (y-axis)), as a function of the voltage on gate 2 ($V_2$) that controls the relative alignment of the two levels [measured data represented in Fig.~\ref{fig:figS4}(c)]. Making $V_2$ more positive the tunnel rate will increase, due to the participation of the excited state in the tunneling process. Each tunnel rate is extracted by recording and processing on the fly (using an FPGA) the tunnel events during the initialization stage ($\sim$5 ms) and fitting an exponential to the histogram of the number of events versus time $\tau$ [Fig.~\ref{fig:figS4}(a)].  Here $\tau$ is the time between the start of the initialization stage and the actual tunnel-in event.
For the experiment we used $V_2$ around -196 mV [red points in Fig.~\ref{fig:figS4}(c)], which has been optimized for a good readout and initialization process. As schematically represented inside the orange frame of Fig.~\ref{fig:figS4}(b), if we model the electric noise from the environment as a relative fluctuation of $V_2$ ($\delta V_2$), we can write $\delta \Gamma_\text{IN}\sim \frac{\partial \Gamma_\text{IN}}{\partial V_2}\delta V_2$. In this way we use $\delta \Gamma_\text{IN}$, the tunnel rate fluctuations during the initialization stage (of the 4-stages pulse sequence), as a probe of electric noise.
(ii) The same 4-stage voltage pulse scheme allows us to perform a rotation of the electron spin and to read it out in the last stage. Due to the coupling with the environmental noise, the spin Larmor frequency, $f_0^{(5)}$, fluctuates [see blue frame of Fig.~\ref{fig:figS4}(b)] over time, as reported in the CW measurement over time in Fig.~\ref{fig:figS4}(d). We fit each resonance trace to a Gaussian in order to extract the center of it (Larmor frequency).
(iii) We repeat this measurement over time (each trace takes tens of seconds), in order to get enough statistics, and plot, as reported in Fig.~3(d), the extracted $\Gamma_\text{IN}$ ($x$ axis) and Larmor frequency fluctuation $\Delta f_0$ ($y$ axis) traces, for each measurement cycle. 

\begin{figure}
\includegraphics[width=5.5cm]{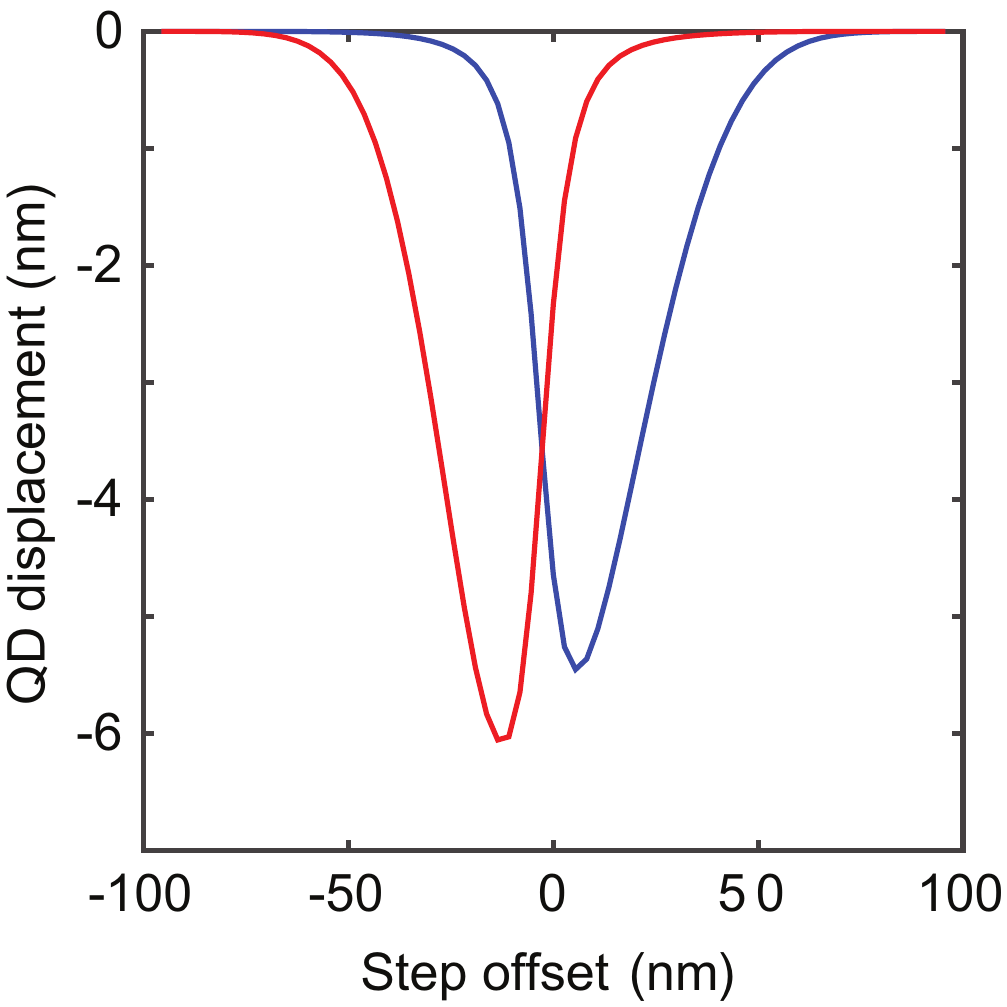}
\caption{Tight binding simulations of the displacement of the electron center of mass for the ground (blue) and the first excited (red) valley states as a function of the position of an atomic step at the quantum well interface, relative to the quantum dot. The geometry for the simulation is shown in Fig.~3(b) of the main text, and the simulation parameters are also given there. 
}
\label{fig:FigS1b}
\end{figure}

A more rigorous estimate of the correlation in time between $\Gamma_\text{IN}(t)$ and $\Delta f_0(t)$, evaluated according to the Pearson product-moment correlation coefficient~\cite{NoteS7} gives a value $\sim-$0.5 indicating a modest correlation in time.\\
The sensitivity of this procedure is related to the ability to distinguish small charge fluctuations and it is a function of the static gate voltage configuration ($V_2$) chosen for the measurement. In fact, from Fig.~\ref{fig:figS4}(c) we can clearly notice that we are not working yet in the most sensitive gate voltage configuration (maximal $\partial \Gamma_\text{IN}/\partial V_2$). Instead, the specific electrostatic configuration has been chosen in order to optimize the spin initialization-readout.

\section{S.IV. Simulation of valley-orbit splitting versus dot position} \label{sec:simulations}

The simulation reported in Fig.~3(c) of the main text represents the result of a 2D tight binding calculation of valley splitting in the presence of interfacial disorder, in the form of a single atomic step. The fact that the simulation is 2D rather than 3D affects just the magnitude of $E_{01}$, but not its qualitative dependence on the step offset. The quantum well barrier has been chosen to be 160 meV, corresponding to 30$\%$ Ge. For simplicity, we assumed a parabolic confinement potential for the dot of size $\sqrt{ \langle x^2 \rangle } = 21.1$ nm, corresponding to an orbital energy splitting of $\hbar \omega = 0.45$ meV \cite{Kawakami2014}. The electric field and the quantum well width have been chosen to be $1.5\times 10^6$ V/m and 13 nm, respectively. The experimental quantum well is nominally 12 nm and the experimental electric field is not well known.

There are 3 competing effects that determine the energy splitting between the two lowest eigenstates, which we usually refer as `valley splitting', for simplicity. (i) The first is the actual valley splitting, which is maximized when the electron sees no step. For the electric field assumed here, the valley splitting can vary by about 75~$\mu$eV, as shown in Fig.~3(c) of the main text, depending on the position of the step relative to the dot. In Fig.~S7, we see that the ground and excited states are shifted to the left or right from their unperturbed positions, due to the step. Moreover, they are shifted differently, as a signature of valley-orbit coupling. (ii) The second effect is related to the electrostatic energy. When the electron sits to the left of the step, it gains a little energy because the energy is given by $eEz$, where $z$ is the vertical position. For a single atom step with $E = 1.5\times10^6$ V/m, this energy scale is 200~$\mu$eV, making it comparable to the valley splitting. This competition between the electrostatic and valley splitting perturbations caused by a step is always present, for any value of the electric field, because the electrostatic energy and the valley splitting both have a linear dependence on $E$. (iii) The third competing effect is the confinement energy of the parabolic dot: when the electron center of mass moves to the left or right due to the step, this increases the average energy of the electron in the parabola. As shown in Fig.~S7, the maximum displacement of the dot is 5-6~nm, when the step is positioned 10-20~nm from the center of the parabola. This corresponds to an energy shift of about 10~$\mu$eV, which is slightly lower than the maximum valley splitting and electrostatic energy scales. However, for a step offset of 20~nm, the valley splitting and electrostatic energy scales are also somewhat smaller, resulting in a competition between all three energies. This competition causes the left-right symmetry of Figs.~2(c) and S7 to be broken. The competition also makes it difficult to gain much intuition about the final results. Without the competing effects, we would expect the valley splitting to go to zero as the ground state changes from `even' to `odd' in the valley parameter. But the electron always wants to maximize the energy splitting between the two eigenstates, and the competing effects (in particular, the shifting of the electrons to the left and right) allow it to keep a nonzero energy splitting at all times. It is interesting that this minimum valley splitting ($\sim$25 $\mu$eV) is very close to the one we observed in our experiments.


\begin{thebibliography}{37}
\bibitem{NoteS1}
In typical resonant driving experiments, the LZ process is
  actually more general than what is shown in the figure, because it can occur
  at any $B$-field. The process indicated in the figure implies an orbital
  excitation combined with a spin flip, which we discuss below.
\bibitem{NoteS2}
Note that the standard theory assumes the qubit is driven
  all the way through the energy level anticrossing in Eq.~(\ref {eq:H});
  however, the interference effects observed in our simulations occur even
  without passing through the anticrossing.
\bibitem{NoteS3}
The simulations are dimensionless and the corresponding
  energy splitting is given by $E_{01}=\protect \sqrt {\epsilon _0^2+\Delta
  ^2}=\protect \sqrt {2}$.
\bibitem{NoteS4}
The charge noise has a $1/f$ dependence, therefore it is a
  relevant noise contribution.
\bibitem{NoteS5}
Changing the gate voltage also will result in a changing of
  QD position in the micromagnet magnetic field gradient but the derivative of
  the gradient with respect to the position is quite small as we can estimate
  from simulation and measure from the red data in Fig.~3(a).
\bibitem{NoteS6}
 Furthermore, assuming that the linewidth of the inter-valley
  spin-flip resonance ($\sigma _{f_0^{(5)}}=\protect \sqrt {2}\pi /T_2^\ast
  =2.1$ MHz) is determined by the valley energy splitting fluctuation caused by
  charge noise, by using the measured $\Delta f_0^{(5)}/\Delta V_2^{p}=9.2$
  MHz/mV we can translate it into an equivalent standard deviation of gate
  voltage noise (for the specific gate 2 used to perform the measurement)
  $\sigma _{V_2}\sim 230$ $\mu $eV.
\bibitem{NoteS7}
It represents a measure of the linear correlation
  (dependence) between two variables $X$ and $Y$, giving a value between +1 and -1
  inclusive, where +1 stands for total positive correlation, 0 no correlation,
  and -1 total negative correlation.
\end{thebibliography}
\end{document}